\documentclass[%
 aip,
 amsmath,amssymb,
 reprint,%
]{revtex4-2}

\usepackage{graphicx}
\usepackage{dcolumn}
\usepackage{bm}

\usepackage[utf8]{inputenc}
\usepackage[T1]{fontenc}
\usepackage{mathptmx}
\DeclareMathAlphabet{\mathcal}{OMS}{cmsy}{m}{n}
\usepackage{etoolbox}
\usepackage{hyperref}
\usepackage{braket}
\usepackage{caption}
\usepackage{subcaption}
\usepackage{bbold}
\usepackage{ulem}
\usepackage{amssymb}
\usepackage[dvipsnames]{xcolor}
\usepackage[symbol]{footmisc}

\usepackage{tikz}
\usetikzlibrary{shapes.geometric, arrows}
\usetikzlibrary{quantikz}
\usepackage[outline]{contour}

\definecolor{qiskit-purple}{RGB}{97, 48, 189}
\definecolor{qiskit-purple-light}{RGB}{174, 125, 255}

\makeatletter
\def\@email#1#2{%
 \endgroup
 \patchcmd{\titleblock@produce}
  {\frontmatter@RRAPformat}
  {\frontmatter@RRAPformat{\produce@RRAP{*#1\href{mailto:#2}{#2}}}\frontmatter@RRAPformat}
  {}{}
}

\def\frontmatter@thefootnote{%
 \altaffilletter@sw{\@fnsymbol}{\@fnsymbol}{\csname c@\@mpfn\endcsname}%
}%

\DefineFNsymbolsTM{otherfnsymbols}{%
  \textdagger    \dagger
  \textasteriskcentered *
  \textdaggerdbl \ddagger
  \textsection   \mathsection
  \textbardbl    \|%
  \textparagraph \mathparagraph
}%

\setfnsymbol{otherfnsymbols}

\makeatother

\begin{document}

\preprint{AIP/123-QED}

\title{Quantum Simulation on Noisy Superconducting Quantum Computers}
\author{Kaelyn J. Ferris}
\thanks{Authors contributed equally to this work}
\affiliation{IBM Quantum, IBM T.J. Watson Research Center, Yorktown Heights, NY 10598, USA}
 \affiliation{Department of Physics \& Astronomy, Ohio University, Athens, Ohio 45701, USA}

\author{A.J. Rasmusson}%
\thanks{Authors contributed equally to this work}
\affiliation{IBM Quantum, IBM T.J. Watson Research Center, Yorktown Heights, NY 10598, USA}
\affiliation{Indiana University Department of Physics, Bloomington, Indiana 47405, USA}

\author{Nicholas T. Bronn}
\affiliation{IBM Quantum, IBM T.J. Watson Research Center, Yorktown Heights, NY 10598, USA}

\author{Olivia Lanes}
\thanks{email: olivia.lanes@ibm.com}
\affiliation{IBM Quantum, IBM T.J. Watson Research Center, Yorktown Heights, NY 10598, USA}

\date{\today}

\begin{abstract}
Quantum simulation is a potentially powerful application of quantum computing, holding the promise to be able to emulate interesting quantum systems beyond the reach of classical computing methods. Despite such promising applications, and the increase in active research, there is little introductory literature or demonstrations of the topic at a graduate or undergraduate student level. This artificially raises the barrier to entry into the field which already has a limited workforce, both in academia and industry. Here we present an introduction to simulating quantum systems, starting with a chosen Hamiltonian, overviewing state preparation and evolution, and discussing measurement methods.  We provide an example simulation by measuring the state dynamics of a tight-binding model with disorder by time evolution using the Suzuki-Trotter decomposition.  Furthermore, error mitigation and noise reduction are essential to executing quantum algorithms on currently available noisy quantum computers. We discuss and demonstrate various error mitigation and circuit optimization techniques that significantly improve performance. All source code is freely available, and we encourage the reader to build upon it\footnote{\url{https://github.com/qiskit-research/qiskit-research/tree/main/docs/tutorial_quantum_sim}}. 
\end{abstract}

\maketitle

\section{\label{sec:intro}Introduction}



When simulating quantum systems, the needed computational resources grow exponentially with the system size due to the exponential growth of the Hilbert space. This quickly outgrows classical computational resources, leaving many interesting quantum systems out of reach.  There are, however, a variety of classical approaches which approximate quantum systems such as quantum Monte-Carlo methods \cite{foulkes2001quantum}, tensor network methods \cite{Augusiak2012,montangero2018intro,orus2019tensor}, density functional theory~\cite{sholl2011density}, and dynamical mean field theory~\cite{Vollhardt2012DMFT}. The accuracy of these approximations is improving, but the full simulation of highly entangled, many-body quantum systems remains a fundamental challenge.

Simulating quantum systems on an analog or digital quantum computing platform~\cite{divincenzo2000physical} could sidestep the exponential issue altogether since the Hilbert space of a quantum computer could be made to grow exponentially alongside the Hilbert space of the system of interest. However, achieving such a computational advantage is not guaranteed \cite{lee2022there} and, in general, is a hard problem.  Discovering and demonstrating efficient quantum methods for simulating quantum systems is an ongoing field of research~\cite{georgescu2014quantum, Tacchino2020} with various methods being proposed regularly such as variational algorithms \cite{cerezo2021variational,tilly2021variational} or digitized Hamiltonian evolution \cite{Suzuki90, Suzuki91, lanyon2011universal, barends2015digital, salathe2015digital}. Choosing which technique to use greatly depends on the system of interest, the desired initial state, the observables of interest, and what experimental hardware will execute the simulation.

Quantum simulation on currently available noisy quantum computers \cite{preskill2018quantum} is a recent accomplishment~\cite{cade-swapnetworks,Karamlou2022-QuantumWalks}. As such, standardized simulation techniques are not widely established. This new frontier offers an opportunity for newcomers, such as graduate student researchers, to contribute, but keeping up with recent developments can be a hurdle due to the field's fast-paced nature. Compounding this issue, there is a growing need for a quantum educated workforce with experience in state-of-the-art methods such as quantum simulation \cite{aiello2021achieving}.

\begin{figure*}[t!]
\begin{tikzpicture}[node distance=6cm]
\tikzstyle{process} = [rectangle, rounded corners, text width=4.5cm, minimum width=5.0cm, minimum height=2.5cm, draw=black, text centered, fill=qiskit-purple-light, font=\normalsize]
\tikzstyle{process2} = [rectangle, rounded corners, text width=7.5cm, minimum width=8.0cm, minimum height=2.5cm, draw=black, text centered, fill=qiskit-purple-light, font=\normalsize]

\tikzstyle{footnote} = [text width=4.5cm, minimum width=5.0cm, minimum height=2.5cm, draw=black, text centered]
\tikzstyle{footnote2} = [text width=7.5cm, minimum width=8.0cm, minimum height=2.5cm, text centered]

\tikzstyle{arrow} = [thick,->,>=stealth]

\node (one) [process, text=black] {\normalsize{Identify system Hamiltonian} \\ \textcolor{qiskit-purple-light}{a} \\ { {\large $H = \sum_i h_i$}}};
\node[footnote, draw=none, align=left, below left=-0.4cm and -4.95cm of one] {{\bf Fermionic}\\Bosonic\\Spin\\Mixed};

\node (two) [process, right of=one, text=black] {\normalsize{Hamiltonian encoding} \\ \textcolor{qiskit-purple-light}{a} {\includegraphics[width=1.\textwidth]{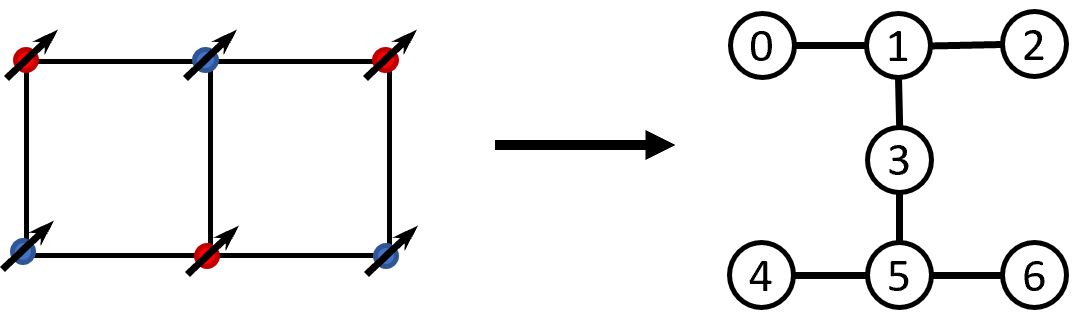}}};
\node[footnote, draw=none, align=left, below left=-0.4cm and -4.95cm of two]
{\textbf{Jordan-Wigner}\cite{Jordan1928}\\Bravyi-Kitaev\cite{Bravyi2002}\\Parity\cite{Seeley2012}\\Higher order spin \cite{mathis2020toward}};

\node (three) [process,  right of=two, text=black] {\normalsize{State preparation} \\ \textcolor{qiskit-purple-light}{a} \\ {\includegraphics[width=0.8\textwidth]{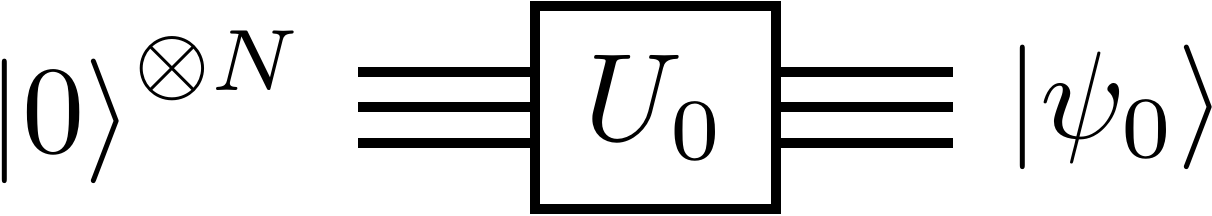}}};
\node[footnote, draw=none, align=left, below left=-0.62cm and -4.95cm of three]
{{\textbf{Excitation in Fock basis}}\\  Maximally mixed state \\ Slater determinant~\cite{Wecker-givensrotation,Sung2022}}; 
\node (four) [process2,  below right=2.1cm and -5.0cm of one, text=black] {\normalsize{State evolution and measurement} \\ \textcolor{qiskit-purple-light}{a} \\ {\includegraphics[width=0.9\textwidth]{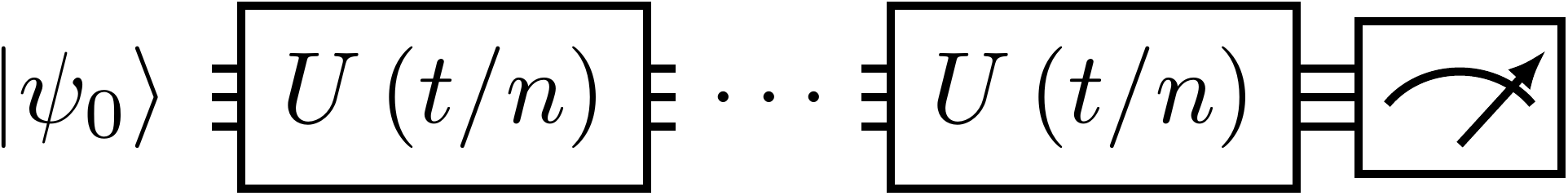}}};
\node[footnote2, draw=none, align=left, below left=-0.62cm and -7.9cm of four] {{\bf Suzuki-Trotter}\cite{Suzuki90,Suzuki91}\\Linear combination of \\ \quad unitaries\cite{Childs-LCU}};
\node[footnote2, draw=none, align=left, below right=-0.22cm and -3.75cm of four] {{\bf Direct measurement} \\ Correlation function \cite{Ortiz2001, Somma2002} \\ Phase estimation \cite{kitaev1995quantum, Somma2002, Corcoles2021} \\ Spectroscopic eigenvalue \\ \quad estimation\cite{Stenger2022}};

\node (five) [process2,  below left=2.1cm and -5.0cm of three] {\normalsize{Error mitigation} \\ \textcolor{qiskit-purple-light}{a} \\
{\includegraphics[width=1.\textwidth]{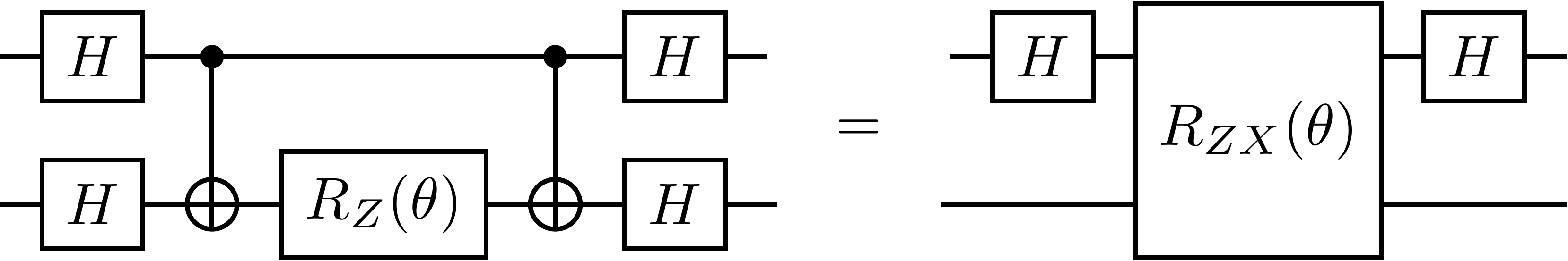}}};
\node[footnote2, draw=none, align=left, below left=-0.2cm and -7.9cm of five] {{\bf Minimize circuit depth}\\{\bf Measurement error \\ \quad mitigation} \cite{nation2021scalable}\\Probabilistic error correction\cite{Temme2017}\\Pauli twirling \cite{bennett1996mixed}};
\node[footnote2, draw=none, align=left, below right=-0.25cm and -3.75cm of five] {{\bf Optimize qubit topology}\cite{mapomatic}\\{\bf Scaled native gates}\cite{earnest2021pulse}\\{\bf Post-selection \cite{stanisic2021observing, noel2022measurement, Sung2022} } \\Zero noise extrapolation \cite{Temme2017, kandala2019error, Kim2021} \\Dynamical decoupling \cite{viola1999dynamical, jurcevic2021demonstration,  ezzell2022dynamical}};

\draw [arrow] (one) -- (two);
\draw [arrow] (two) -- (three);
\draw [arrow] (four) -- (five);
\draw[arrow] (three.east) -- ++(0.25,0) -- ++(0,-3) -|  (four);
\end{tikzpicture}
\caption{Conceptual workflow for constructing and executing a quantum simulation algorithm on a quantum computer.  We begin with a physical model we wish to simulate and encode the operators to native qubit interactions and encode the basis states to the qubits' computational basis. Next, an initial state is prepared to begin the simulation. A method of unitary evolution is applied followed by measurement of the desired observable. Before executing the circuit on physical hardware, one should consider using error mitigation techniques to minimize errors and maximize the fidelity of measuring the desired observable. Common methods and techniques used in quantum simulation, along with a small subset of representative references, are enumerated below each purple box with bold indicating the processes used for the example model in this work.}
\label{fig:flowchart}
\end{figure*}
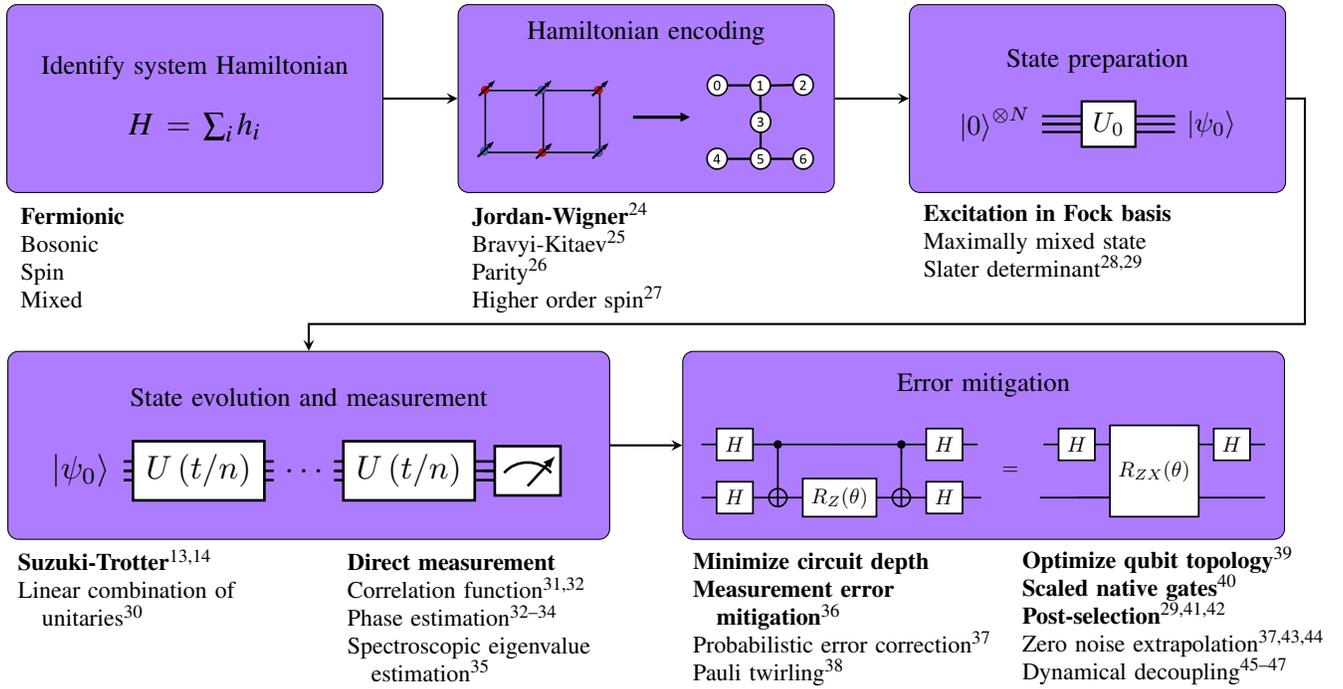

This article addresses these concerns with a graduate student level discussion of the general approach to quantum simulation followed by a minimal working example.  We present a general approach to quantum simulation algorithms in Fig.~\ref{fig:flowchart} and then discuss each step in detail: identifying a system Hamiltonian, transforming to Pauli basis states and operators, preparing an initial state, evolving the state, and measuring the desired observables. For each of these broad topics, we have included a few references and provide further discussion in the appendix sections. After a brief introduction in each topic section, we make specific choices about what techniques to discuss with the goal of demonstrating the dynamics of a tight-binding chain with a defect.  We then present state-of-the-art experimental error mitigation techniques used on current quantum computing devices.  The source code and experimental error mitigation techniques are available in an open-access format so the reader can readily build upon them \cite{tutorial-notebook}.

This article's organization follows the flowchart of Fig. \ref{fig:flowchart}.
We begin by reviewing Pauli and second quantization operators and the Jordan-Wigner (JW) transform in Sec.~\ref{sec:ham-encoding}.  Section~\ref{sec:state-prep} discusses state preparation. Section \ref{sec:state-evolution} reviews a common procedure for state evolution, the Suzuki-Trotter decomposition.  Measurement is discussed in Sec.~\ref{sec:measurement}.  We then move to a practical example---a one-dimensional tight-binding model with a defect---and apply these ideas in Sec.~\ref{sec:tb-model} followed by introducing state-of-the-art techniques for reducing significant errors in Sec.~\ref{sec:q-sim-on-real-qc}.  In Sec.~\ref{sec:conclusion}, we conclude by briefly sharing several extensions students can pursue as advanced course projects. All source code is written in Python using Qiskit~\cite{Qiskit}: an open-source platform for quantum computation. The source code is organized in an openly available Jupyter notebook \cite{tutorial-notebook}. All quantum simulations in this article were run on IBM's 7-qubit device $\textit{ibm\_lagos}$.

\section{Hamiltonian encoding for quantum simulation} \label{sec:ham-encoding}

The process of preparing dynamic quantum simulations begins with identifying the Hamiltonian of the physical system we wish to simulate.  Both the operators of this Hamiltonian as well as its associated Hilbert space must be mapped to the accessible operators and Hilbert space of the quantum computer.  We refer to this mapping as ``encoding'' the Hamiltonian.  This encoding highly depends on both the chosen mapping and qubit topology.

Often, a Hamiltonian $H$ with local interactions is selected (e.g. a lattice of atoms which only interact with their nearest neighbors)
\begin{equation}
    H = \sum_i^N h_i
\label{eq:ham-sub-sys}
\end{equation}
where $h_i$ is a Hamiltonian acting on one of $N$ local subsystems of $H$.
In the simplest case, the encoding leaves all interactions as local, two-qubit gates.


Another important aspect of $H$ for quantum simulation is the type of operators used to construct it. Three types of operators are typically used and can be readily encoded onto a quantum computer: fermionic, bosonic, and spin operators. Fermionic and bosonic operators are often written in second quantization, the language of creation and annihilation operators while spin (fermionic and bosonic spin) operators are expressed in terms of the spin matrices, which for spin-$1/2$, are the Pauli matrices.

Hamiltonians constructed from the Pauli operators can be directly encoded onto a quantum computer. The Pauli operators in the computational basis are represented in matrix form as
\begin{align}
    \sigma^x &=
    \begin{pmatrix}
    0 & 1 \\
    1 & 0
    \end{pmatrix} \\
    \sigma^y &=
    \begin{pmatrix}
    0 & -i \\
    i & 0 
    \end{pmatrix} \\
    \sigma^z &=
    \begin{pmatrix}
    1 & 0 \\
    0 & -1 
    \end{pmatrix}\;
\end{align}
with the commutation relation $\left[\sigma^i, \sigma^j \right] = 2 i \epsilon_{ijk} \sigma^k$ for $i,j,k \in \{x,y,z\}$ and the Levi-Civita tensor $\epsilon_{ijk}$. Pauli operators are also commonly represented by the capital letters $X$, $Y$, and $Z$ for convenience where $X \equiv \sigma^x$, $Y \equiv \sigma^y$, and $Z \equiv \sigma^z$. Spin systems with spin higher than $1/2$ can also be encoded onto a quantum computer\cite{mathis2020toward}.

Hamiltonians constructed from creation and annihilation operators first require a transformation to Pauli operators that respects the statistics of the particles they describe. There are several such transformations for fermions. A common choice is the Jordan-Wigner transformation~\cite{Jordan1928}. The Bravyi-Kitaev~\cite{Bravyi2002} and parity~\cite{Seeley2012} mappings are more recent fermionic encodings employing one qubit per fermion. Bosonic operators can be transformed by using the Holstein-Primakoff transformation~\cite{Holstein1940} or directly mapping the Fock states to a sub-basis of the available bosonic modes~\cite{Somma2003}.  Other encodings, including both bosonic systems and systems with higher spin, are actively being researched~\cite{Sawaya2020}.  The required resources for quantum simulation have become a recent research question~\cite{Steudtner2017}, and more efficient encodings for certain Hamiltonian classes such as those that preserve particle number continue to be explored~\cite{Kirby2021}.

\subsection{Operators in second quantization}
The second quantization formalism is primarily used in quantum many-body systems to describe states associated with an ensemble of particles. The wavefunctions for these systems are typically described as a Fock state, often written as 
\begin{equation}
    \ket{\psi} = \ket{n_0, n_1, n_2,...,n_i,...}
\end{equation}
where $n_i$ denotes the number of particles in the $i^{th}$ one-particle state. The fundamental operators of this formalism are the creation and annihilation operators $\hat{a}^\dagger$ and $\hat{a}$ respectively. These operators act on each of the one-particle states, and their eigenvalues are related to the square root of the particle number $n_i$ by
\begin{align}
    \hat{a}^\dagger_i\ket{n_0, n_1,...,n_i,...} &= \sqrt{n_i+1}\ket{n_0, n_1,...,n_i+1,...} \\
    \hat{a}_i\ket{n_0, n_1,...,n_i,...}& = \sqrt{n_i}\ket{n_0, n_1,...,n_i-1,...} \;.
\end{align}

The commutation relations between $\hat{a}^{\dagger}$ and $\hat{a}$ differ depending on the particle type: bosons ($\hat{b}^\dagger, \hat{b}$) or fermions ($\hat{c}^\dagger, \hat{c}$). For bosonic systems, these operators have commutation relations of the form
\begin{align}
    &[\hat{b}_i,\hat{b}_j^\dagger] = \delta_{ij} \\
    &[\hat{b}_i, \hat{b}_j] = [\hat{b_i}^\dagger, \hat{b}_j^\dagger] = 0
    \;,
\end{align}
and for fermionic systems, the commutator is replaced with the anticommutator
\begin{align}
    &\{\hat{c}_i,\hat{c}_j^\dagger\} = \delta_{ij} \\
    &\{\hat{c}_i, \hat{c}_j\} = \{\hat{c}_i^\dagger, \hat{c}_j^\dagger\} = 0 \;.
\end{align}
Additionally, each $n_i$ in the Fock basis of a fermionic system is restricted to $n_i=\left\{0,1\right\}$ due to the Pauli exclusion principle whereas for a bosonic system $n_i=\left\{0,1,2,\ldots \right\}$.

Another operator common to this formalism is the number operator
\begin{equation}
    \hat{n}_i = \hat{a}_i^\dagger\hat{a}_i
\end{equation}
which leaves a given state $\ket{\psi}$ unchanged but returns the number of particles $n_i$ as an eigenvalue
\begin{equation}
    \hat{n}_i\ket{n_0, n_1,\cdots,n_i,\cdots} = n_i\ket{n_0, n_1,\cdots,n_i,\cdots} \;.
\end{equation}


\subsection{Transform to Pauli operators}
In this manuscript, we focus on the Jordan-Wigner transformation~\cite{Jordan1928} (JWT) to map fermionic operators to Pauli operators as it is one of the most common mapping techniques found in the literature. This transformation is built from the qubit (Pauli) raising and lowering operators
\begin{align}
    \sigma^+ &= \frac{\sigma^x-i\sigma^y}{2} \label{eq:c-dagger} \\
    \sigma^- & = \frac{\sigma^x+i\sigma^y}{2} \label{eq:c}
\end{align}
in the computational basis
\begin{align}
    \ket{0} &= \begin{pmatrix}
                1 \\
                0
                \end{pmatrix} \\
    \ket{1} &= \begin{pmatrix}
                0 \\
                1
                \end{pmatrix}       \;.  
\end{align}
Note that the definitions of $\sigma^\pm$ are the opposite of spin-$1/2$ raising and lowering operators. For fermions, note that the anticommutation relation is not preserved $\{\sigma^-,\sigma^+\} \neq 0$, hence the fermionic mapping must be slightly modified. This can be fixed by inserting $Z$ operators such that the transformed operators take the form
\begin{align}
    \hat{c}_i^\dagger &= Z_0 \cdots Z_{i-1} \left( \frac{X_i - iY_i}{2} \right) I_{i+1} \cdots I_N \\
    \hat{c}_i &= Z_0 \cdots Z_{i-1} \left( \frac{X_i + iY_i}{2} \right) I_{i+1} \cdots I_N
    \label{eq:jw-c}
\end{align}
for a system with $N$ fermionic modes. Remember that we have introduced the notation $X_i \equiv \sigma_i^x$, $Y_i \equiv \sigma_i^y$, and $Z_i \equiv \sigma_i^z$ for simplicity and may suppress qubit number in the future because it is indicated by position in the {\it Pauli string}, an ordered tensor product of Pauli operators with the product symbol ($\otimes$) omitted.

\section{State Preparation}
\label{sec:state-prep}
Having encoded the desired Hamiltonian into the quantum computer's gate set and computational basis, the next step is to prepare an initial state.  This preparation is dependent on both the Hamiltonian and the desired observable.

The preparation of an arbitrary n-qubit state
\begin{equation}
    \ket{\psi} = \sum_{i=0}^{2^n-1}a_i\ket{c_i},
\end{equation}
where $a_i \in \mathbb{C}$, $\ket{c_i}$ are the computational basis states, and the normalization condition
\[
\sum_{i=0}^{2^n-1}|a_i|^2 = 1,
\]
can generally be approached by decomposing an operator $\hat{O}$ which takes an initial state $\ket{0}^{\otimes n}$ to the desired state $\ket{\psi}$ such that $\hat{O}\ket{0}^{\otimes n} = \ket{\psi}$.  In general this requires an exponential number of CNOT gates of $\mathcal{O}(2^n)$~\cite{Plesch-stateprep}.  Including auxiliary qubits, the required depth can be reduced to $\mathcal{O}(n)$ at an exponential cost of extra qubits $\mathcal{O}(2^n)$~\cite{Zhang-ancilla-stateprep}.  In general, this is a hard problem and an active area of research~\cite{Rancic2021, Jian-gaussianStates}.  

For many simulation algorithms, all that is required is some amount of overlap between the initial state and the eigenstates of the Hamiltonian.  Given an observable $A$ with eigenvalues $a_i$ and eigenstates $\ket{\varphi_i}$, one can write an arbitrary state $\ket{\psi}$ as a linear superposition of the eigenstates of $A$
\begin{equation}
    \ket{\psi} = \sum_i c_i \ket{\varphi_i}.
\end{equation}
Assuming we are able to prepare the state $\ket{\psi}$, the probability of measuring an eigenvalue $a_i$ is the amount of overlap between $\ket{\psi}$ and $\ket{\varphi_i}$
\begin{align}
    \bra{\psi}A\ket{\psi} =& \sum_i |c_i|^2\bra{\varphi_i}A\ket{\varphi_i} \nonumber \\
    =& \sum_i|c_i|^2 a_i.
\end{align}
Extracting the eigenvalues $a_i$ requires additional post-processing or measurement techniques which we will discuss in  Section~\ref{sec:measurement}.  Often times a good choice of an initial state is either the state of maximal superposition or the maximally mixed state.  This ensures that there is at least some overlap over all of the eigenstates of a Hamiltonian.

In other cases, we may be interested in preparing an eigenstate of a Hamiltonian explicitly.  In fermionic quadratic Hamiltonians, for example, eigenstates (including Slater determinants) can be prepared efficiently~\cite{verstraete-ferminoicstates, Jian-gaussianStates, Wecker-givensrotation}.  These states are prepared using a combination of Givens rotations and particle-hole transformations and are summarized in Appendix~\ref{appendix:state-prep} and~\ref{appendix:gaussian-state}.

\section{State Evolution} \label{sec:state-evolution}

Once a method for state preparation has been identified, we are ready to decide the best method for evolving the initial state. There are a variety of approaches such as hybrid variational algorithms---namely the the Hamiltonian Variational Ansatz~\cite{cade-swapnetworks,cubitt2018universal}, quantum walks~\cite{Karamlou2022-QuantumWalks}, or quantum signal processing~\cite{Somma2002,Somma2019,Griffiths-QFFT}. Selecting which method to use depends on the observable that will be measured.

In this article, we will focus on algorithms which directly implement the time evolution operator $U(t)=e^{-iHt}$ or an approximation thereof.  In general, one may decompose any unitary into single and two-qubit gates, but it is often inefficient to do so.  We summarize the approach to approximating the time evolution operator via the Suzuki-Trotter decomposition below.  Other methods, such as the method of Linear Combination of Unitaries~\cite{Childs-LCU} can also be used to approximate $U(t)$, which is summarized in Appendix~\ref{appendix:LCU}.

\subsection{Unitary time evolution}

Quantum dynamics such as response and time-correlation functions are often used in order to characterize quantum systems.  Dynamics are typically obtained via the time evolution of a quantum state $\ket{\psi(t)}$ for time $t$ under a time-independent Hamiltonian $H$.  This can be described by the solution to the time dependent Schr\"{o}dinger equation (setting $\hbar = 1$)
\begin{equation}
    |\psi(t)\rangle = e^{-i t H}|\psi(0)\rangle \;
\end{equation}
where we'll denote $e^{-itH}=U(t)$ for simplicity. $U(t)$ is most likely not in a form we can directly input to qubits, particularly if the Hamiltonian encoding leaves the subsystems ($h_i$) as non-local interactions. Ultimately, $U(t)$ needs to be decomposed into a product of operators which are native to the quantum computer, and different quantum computing platforms have different native operations.

Note that for noncommuting terms in $H$ the unitary time evolution operator cannot be directly decomposed into a product  $e^{-it\sum_i h_i} \neq \prod_i e^{-ith_i}$.  The consequence is that there is no simple decomposition from the exponentiated sum of local interactions to a product of quantum gates. However, approximation methods can be employed to do such a decomposition (described below).  Additionally, one may take advantage of the specific commutators of each $h_i$ or the fact that some subsets with $H$ possess identities or relations which can simplify the implementation of $U(t)$.~\cite{alg-compression}

\subsubsection{Trotterization}
The Suzuki-Trotter (ST) decomposition~\cite{Suzuki90,Suzuki91} is a common method for approximately decomposing time evolution operators. After decomposition, the approximate unitary can be readily modified to match the native interactions and gate operations of any quantum computer.

The ST decomposition approximates the sum of exponentiated operators into a product of operators in the limit of small times
\begin{equation}
    U(t) = e^{-i t\sum_{i=1}^L h_i} \approx \left[\prod_{i=1}^L e^{-i h_i dt}\right]^m = U_1\left(dt\right)^m
\label{eq:Suzuki-Trotter-decomp}
\end{equation}
where $m$ is the number of Trotter ``steps'' and $dt=t/m$. The error of this decomposition scales as $\mathcal{O}(dt^2)$. Equation \eqref{eq:Suzuki-Trotter-decomp} is not a unique approximation as it may be taken out to higher orders of $dt$. For example, there is a second order decomposition
\begin{equation}
    U_2\left(dt\right) = \left[U_1(dt/2) U_1(dt/2)^T \right]^m
\end{equation}

\noindent
where $U_1(dt/2)^T$ denotes the product of Eq.~\eqref{eq:Suzuki-Trotter-decomp} in reverse order, with error scaling as $\mathcal{O}(dt^3)$. Higher order decompositions are defined by the recurrence relation
\begin{equation}
    U_{2k}\left(dt\right) = U_{2k-2}\left(p_kdt\right)^2U_{2k-2}\left((1-4p_k)dt\right)U_{2k-2}\left(p_kdt\right)^2
\end{equation}
which is defined over even orders of $k$ since $U_{2k}=U_{2k-1}$~\cite{Suzuki91}. The coefficients $p_k$ are defined as
\begin{equation}
    p_k\equiv \frac{1}{4-4^{\frac{1}{2k-1}}} \;.
\end{equation}
The accuracy of the Suzuki-Trotter approximation is then determined by the order $k$ of the decomposition and the number of Trotter steps $m$ (where more steps leads to higher approximation accuracy at the cost of increased circuit depth).

\section{Measurement} \label{sec:measurement}

The final part of a quantum simulation algorithm is the measurement of some quantity of interest.  This may be simply the probability amplitudes $p_i = |\braket{q_i|\psi(t)}|^2$ where $q_i$ are the qubit basis states (e.g. see Section~\ref{sec:tb-model}).  Other approaches such as the direct measurement of an observable~\cite{Georgescu-QSim} are also feasible.  Direct measurements of an observable will directly measure the system qubits, elucidating expectation values such as the total energy or magnetization.

If an observable $\mathcal{O}$ can be transformed to some Pauli operator $\mathcal{P}$ so that expectation values have the form
\[
\langle \mathcal{O} \rangle = \langle \psi(t) |\mathcal{P}|\psi(t)\rangle,
\]
we can calculate $\langle \mathcal{O}\rangle$ by processing the count data returned from the quantum processor.  Commuting observables may be grouped together to reduce the effective number of experiments needed.  In most quantum computing systems the standard measurement is in the $Z$-basis, but other Pauli operators can be measured by an appropriate rotation into the $X$ or $Y$ bases prior to measurement.  


In practice, every expectation value comes with an uncertainty that depends on quantum projection noise and finite sampling. Quantum projection noise comes from the random outcomes quantum superposition (or entangled) states give when measured in the computation basis. The standard deviation of which follows $\sqrt{p(1-p)}$ where $p$ is the probability of measuring the state $\ket{1}$ (or $\ket{0}$). Given $N$ repeated measurements, the effect of finite sampling can also be included in the uncertainty leaving the total standard deviation of a quantum expectation value to be
\begin{equation}
    \sigma = \sqrt{\dfrac{p(1-p)}{N}} \;.
    \label{eq:shot-noise}
\end{equation}
From Eq. \eqref{eq:shot-noise}, we can estimate how many repeated measurements are needed to achieve a particular accuracy.

Additionally, more sophisticated measurements and post-processing methods can allow us to probe the properties of a system such as correlation functions~\cite{Ortiz2001} or the eigenvalue spectrum~\cite{Somma2019, Stenger2022}.  We summarize a few of these algorithms in Appendix~\ref{appendix:aux-measurement},~\ref{appendix:eig-estimation}, and~\ref{appendix:spectroscopic-measurement}.

Having discussed the recipe for quantum simulation of state dynamics and the various ingredients available to us, we now demonstrate each of the ideas discussed in the above sections for a tight-binding model, both in this manuscript and in the associated Jupyter notebook \cite{tutorial-notebook}. Afterward, we introduce increasingly common experimental considerations that mitigate noise and give significant performance improvements.


\section{Example: A tight-binding model}
\label{sec:tb-model}

\begin{figure}[t]
    \centering
    \begin{tikzpicture}[node distance={12mm}, very thick, main/.style = {draw, circle, color=qiskit-purple}, color=black]
    \node[main] (0) {$0$};
    \node[main] (1) [right of=0] {$1$};
    \node[main] (2) [right of=1] {$2$};
    \node[main] (3) [right of=2] {$3$};
    \node[main] (4) [right of=3] {$4$};
    \draw[-, color=qiskit-purple] (0) -- node[midway, above, black] {-$\tau$} (1);
    \draw[-, color=qiskit-purple] (1) -- node[midway, above, black] {-$\tau$} (2);
    \draw[-, color=qiskit-purple] (2) -- node[midway, above, black] {-$\tau_d$} (3);
    \draw[-, color=qiskit-purple] (3) -- node[midway, above, black] {-$\tau$} (4);
    \end{tikzpicture}
    \caption{A graph representation of the physical system that will be simulated (Eq. \eqref{eq:tb-wd-ham}). The purple nodes represent lattice sites where a particle could reside. Each node is connected to its neighbors by a weighted edge representing the hopping strength $\tau$ or $\tau_d$ between sites.}
    \label{fig:lattice}
\end{figure}
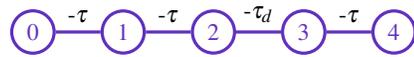

Tight-binding models are a simple starting point towards understanding the electronic band structure of materials \cite{slater1954simplified} and are often used as introductory examples in condensed matter physics. As such, it makes for an excellent introductory model to learn quantum simulation techniques. We apply the theoretical techniques discussed in the previous sections to a 1D tight-binding model with a defect shown in Fig.~\ref{fig:lattice}. Each node in Fig. \ref{fig:lattice} could, for example, be thought of as an atom in a solid with support for a single electron in its orbital. More specifically, the system under study is a five site chain with a hopping strength $\tau$ between all sites except for a defect between sites two and three. The defect is modeled as a change in hopping term from $\tau$ to $\tau_d$ ($\tau_d \neq \tau$), and it disrupts a given particle's wavefunction \cite{vsiber2006dynamics, braumuller2022probing}.  We consider the scenario where a single particle occupies the lattice, and we ignore its spin.
\begin{equation}
H = - \tau \sum_{i \ne 2} \left(c_i^{\dagger} c_{i+1} + c_{i+1}^{\dagger} c_i \right) - \tau_d \left(c_2^{\dagger} c_{3} + c_{3}^{\dagger} c_2 \right) \;.
\label{eq:tb-wd-ham}
\end{equation}

The simplicity of the Hamiltonian serves to focus our attention on the quantum simulation techniques described earlier as well as error mitigation techniques to be discussed later on. Note that this Hamiltonian is a quadratic fermionic Hamiltonian which can be solved exactly \cite{jiang2018quantum, Wecker-givensrotation}. Beyond the physically relevant aspects of this tight-binding model, its purpose in this article is to serve as a full, beginning to end, exposition of quantum simulation (see Fig. \ref{fig:flowchart}). Additionally, it can serve as a launching point for the reader into more complicated models such as Jaynes-Cummings \cite{JC-textbook}, Fermi-Hubbard \cite{hubbard-textbook}, Bose-Hubbard \cite{hubbard-textbook}, or Kitaev spin chain~\cite{kitaev-models-textbook}.


Following the steps of the flow chart in Fig. \ref{fig:flowchart}, we substitute each term in \eqref{eq:tb-wd-ham} with Eq.\eqref{eq:jw-c} to map $H$ to the Pauli operators.  After the JWT, the Hamiltonian simplifies to
\begin{equation}
H_p = - \frac{\tau}{2} \sum_{i \ne 2} \left( X_i X_{i+1} + Y_i Y_{i+1} \right) - \frac{\tau_d}{2}\left( X_2 X_{3} + Y_2 Y_{3} \right) \;.
\label{eq:tb-ham-pauli}
\end{equation}
The qubit states represent the Fock states of the system.
The time evolution operator ($\hbar=1$) is then $U(t) = e^{-i t H_p}$.


\begin{figure}[t]
  \includegraphics[width=0.85\linewidth]{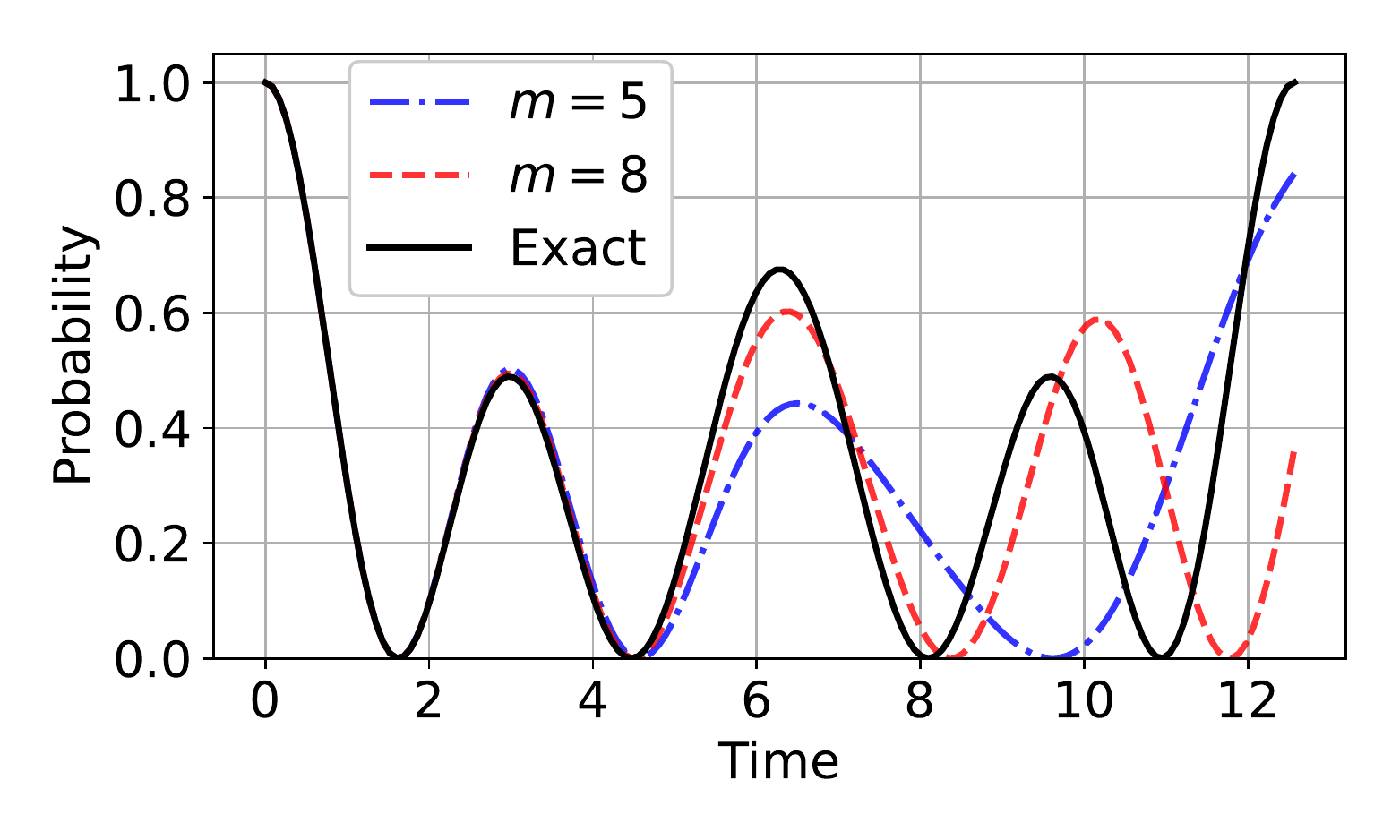} 
  \caption{Classically computed time evolution along site $0$ ($\langle n_0 \rangle$) for various Trotter steps $m$. The evolution from the Trotterized unitary for five (blue dash-dotted line) and eight (red dashed line) Trotter steps do not share this symmetry, but the discrepancy is lessened for the eight trotter step evolution compared to the five step evolution.}
  \label{fig:trotter-error}
\end{figure}

To break the unitary operator into quantum gate operations, the first-order Suzuki-Trotter decomposition from Eq. \eqref{eq:Suzuki-Trotter-decomp} is used. The time operator $U(t)$ for $m$ Trotter steps is approximated to be
\begin{align}
    U_T(t) =& \bigg[\prod_{i \ne 2} e^{\frac{i t}{2m} \tau \left(X_i X_{i+1}\right)} e^{\frac{i t}{2m} \tau \left(Y_i Y_{i+1}\right)} \nonumber \\  &\quad \times e^{\frac{i t}{2m} \tau_d \left(X_2 X_3\right)} e^{\frac{i t}{2m} \tau_d \left(Y_2 Y_3\right)} \bigg]^m \;.
    \label{eq:tb-u-trotterized}
\end{align}

Before executing Eq. \eqref{eq:tb-u-trotterized} as gate operations on a quantum computer, the expected evolution is computed classically with a single excitation as the initial state $|\psi_0\rangle = |00001\rangle$. Here we use Qiskit's little-endian notation (where the rightmost bit is the least significant digit) to represent a single particle on site $0$ in Fig.~\ref{fig:lattice}. Figure \ref{fig:trotter-error} shows the expectation value of the number operator of the 0th site $\langle n_0 \rangle$, which is computed classically, through the evolution time range $\left[ 0, 4 \pi \right]$. Note the reflection symmetry about $t=2\pi$ in the exact solution (black solid line) where the Suzuki-Trotter approximation is not used. The Suzuki-Trotter approximation (Eq. \eqref{eq:tb-u-trotterized}) for $m=5$ (blue dash-dotted line) and $m=8$ (red dashed line) is then computed. Note the inherent error of the Suzuki-Trotter approximation relative to the exact evolution. The time evolution for $m=5$ and $m=8$ does not have the same reflection symmetry about $t=2 \pi$ as the exact evolution. However, as $m$ increases, the Trotter error reduces which can be seen, in part, by how much closer the red $m=8$ curve matches the exact black curve compared to the blue $m=5$ curve.


\subsection{Pauli gate operations}
The operators of Eq.~\eqref{eq:tb-u-trotterized} should now be expanded in terms of Pauli operators in order to implement them on quantum hardware efficiently.  In general, the exponential of any Pauli string $\mathcal{P} = \bigotimes_{i=0}^n P_i$ can be expanded exactly as
\begin{equation}
    e^{i\theta\mathcal{P}} = \cos(\theta)I^{\otimes n} + i\sin(\theta)\mathcal{P}.
\end{equation}
Applying this to the time evolution operator, it can be shown that $U_T(t)$ can be written as a product of two-qubit rotations
\begin{align}
     e^{-i\theta X_i \otimes X_j/2} &= \cos(\theta/2) I_i \otimes I_j - i \sin(\theta/2) X_i\otimes X_j \\
     &\equiv R_{X_i X_j}(\theta) \nonumber \\
     e^{-i \theta Y_i \otimes Y_j/2} &= \cos(\theta/2) I_i \otimes I_j - i \sin(\theta/2) Y_i\otimes Y_j\\
     &\equiv R_{Y_i Y_j}(\theta) \nonumber
\end{align}
where $i$ and $j$ are the qubits the gate operates on and the tensor product $\otimes$ is explicitly included. Depending on the underlying hardware, a quantum computer may apply certain single or two-qubit gates with lower error than other gates. Experimental considerations such as these are important for near-term devices and can greatly impact the simulation results as will be seen in the next section.

In this section, we have identified a Hamiltonian of interest, transformed its operators to Pauli operators, approximately decomposed the time evolution operator into a product of Pauli operators, and identified the relevant quantum gates. The product of quantum gate operations (Eq. \eqref{eq:tb-u-trotterized}) is the quantum circuit that will be executed on the quantum computer. With this in hand, the next task is to optimize the circuit for the specific quantum technology that will be running the quantum simulation.

\section{Quantum simulation on real quantum computers} \label{sec:q-sim-on-real-qc}
Up until this point, everything discussed in this article has been building up to a quantum circuit which simulates state dynamics, but the discussion has mostly ignored the experimental challenge of executing the circuit with minimal noise. How these ideas are executed on real intermediate-scale noisy quantum hardware\cite{preskill2018quantum} can look quite different from how they are introduced algorithmically. To get the best possible performance out of these machines, we can apply a variety of techniques which reduce inevitible errors without the overhead of error correction~\cite{Temme2017, kandala2019error, Kim2021}. In the following section, we will introduce and demonstrate~\cite{tutorial-notebook} some of these increasingly common error mitigation techniques.

Error mitigation aims to reduce errors in a quantum computation without real-time feedback or fault tolerance strategies~\cite{Temme2017}. Error mitigation refers to a collection of methods that suppress noise, often with substantial effect. The techniques listed below are by no means exhaustive, but they serve as a starting point and follow what is quickly becoming standard practice.


\subsection{Optimizing qubit topology}

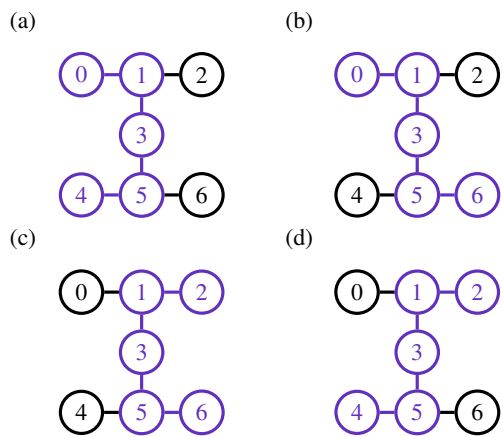
\begin{figure}[b]
    \centering
    \begin{subfigure}[t]{0.2\textwidth}
        \caption{}
        \begin{tikzpicture}[node distance={8mm}, very thick, main/.style = {draw, circle, color=black}, selected/.style = {draw, circle, color=qiskit-purple}, color=qiskit-purple]
        \node[selected] (0) {$0$};
        \node[selected] (1) [right of=0] {$1$};
        \node[main] (2) [right of=1] {$2$};
        \node[selected] (3) [below of=1] {$3$};
        \node[selected] (5) [below of=3] {$5$};
        \node[selected] (4) [left of=5] {$4$};
        \node[main] (6) [right of=5] {$6$};
        \draw[-] (0) -- node[midway, above] {} (1);
        \draw[-, color=black] (1) -- node[midway, above] {} (2);
        \draw[-] (1) -- node[midway, above] {} (3);
        \draw[-] (3) -- node[midway, above] {} (5);
        \draw[-] (4) -- node[midway, above] {} (5);
        \draw[-, color=black] (5) -- node[midway, above] {} (6);
        \end{tikzpicture}
        \label{fig:lagos-ql-choice1}
    \end{subfigure}
    \begin{subfigure}[t]{0.2\textwidth}
        \caption{}
        \begin{tikzpicture}[node distance={8mm}, very thick, main/.style = {draw, circle, color=black}, selected/.style = {draw, circle, color=qiskit-purple}, color=qiskit-purple]
        \node[selected] (0) {$0$};
        \node[selected] (1) [right of=0] {$1$};
        \node[main] (2) [right of=1] {$2$};
        \node[selected] (3) [below of=1] {$3$};
        \node[selected] (5) [below of=3] {$5$};
        \node[main] (4) [left of=5] {$4$};
        \node[selected] (6) [right of=5] {$6$};
        \draw[-] (0) -- node[midway, above] {} (1);
        \draw[-, color=black] (1) -- node[midway, above] {} (2);
        \draw[-] (1) -- node[midway, above] {} (3);
        \draw[-] (3) -- node[midway, above] {} (5);
        \draw[-, color=black] (4) -- node[midway, above] {} (5);
        \draw[-] (5) -- node[midway, above] {} (6);
        \end{tikzpicture}
        \label{fig:lagos-ql-choice2}
    \end{subfigure}
    \begin{subfigure}[b]{0.2\textwidth}
        \caption{}
        \begin{tikzpicture}[node distance={8mm}, very thick, main/.style = {draw, circle, color=black}, selected/.style = {draw, circle, color=qiskit-purple}, color=qiskit-purple]
        \node[main] (0) {$0$};
        \node[selected] (1) [right of=0] {$1$};
        \node[selected] (2) [right of=1] {$2$};
        \node[selected] (3) [below of=1] {$3$};
        \node[selected] (5) [below of=3] {$5$};
        \node[main] (4) [left of=5] {$4$};
        \node[selected] (6) [right of=5] {$6$};
        \draw[-, color=black] (0) -- node[midway, above] {} (1);
        \draw[-] (1) -- node[midway, above] {} (2);
        \draw[-] (1) -- node[midway, above] {} (3);
        \draw[-] (3) -- node[midway, above] {} (5);
        \draw[-, color=black] (4) -- node[midway, above] {} (5);
        \draw[-] (5) -- node[midway, above] {} (6);
        \end{tikzpicture}
        \label{fig:lagos-ql-choice3}
    \end{subfigure}
    \begin{subfigure}[b]{0.2\textwidth}
        \caption{}
        \begin{tikzpicture}[node distance={8mm}, very thick, main/.style = {draw, circle, color=black}, selected/.style = {draw, circle, color=qiskit-purple}, color=qiskit-purple]
        \node[main] (0) {$0$};
        \node[selected] (1) [right of=0] {$1$};
        \node[selected] (2) [right of=1] {$2$};
        \node[selected] (3) [below of=1] {$3$};
        \node[selected] (5) [below of=3] {$5$};
        \node[selected] (4) [left of=5] {$4$};
        \node[main] (6) [right of=5] {$6$};
        \draw[-, color=black] (0) -- node[midway, above] {} (1);
        \draw[-] (1) -- node[midway, above] {} (2);
        \draw[-] (1) -- node[midway, above] {} (3);
        \draw[-] (3) -- node[midway, above] {} (5);
        \draw[-] (4) -- node[midway, above] {} (5);
        \draw[-, color=black] (5) -- node[midway, above] {} (6);
        \end{tikzpicture}
        \label{fig:lagos-ql-choice4}
    \end{subfigure}
    \caption{Qubit layout choices on the seven qubit topology of $\textit{ibm\_lagos}$. The purple nodes in panels (a-d) represent the 5 qubits selected for the tight-binding quantum simulation. The two black nodes in panels (a-d) would not be used. The \texttt{mapomatic} scores for the layouts are: 0.2221 for (a), 0.2144 for (b), 0.2113 for (c), and 0.2191 for (d).}
    \label{fig:lagos-ql-choice}
\end{figure}

Superconducting qubits are typically coupled in a planar topology with nearest-neighbor-type connectivity. If the connectivity of the Hamiltonian does not match the hardware, SWAP operations will be needed to realize the simulation. We call this process routing, and for the tight binding Hamiltonian we've chosen, the connectivity is 1D and can be realized with no SWAPs. In general, this is not always the case, and many SWAPs could be needed. Since SWAPs are composed of 3 CNOT gate operations---and these two-qubit gates are the leading source of error in today's hardware---they can quickly reduce a circuit's performance. Therefore, it is crucial to minimize the number of SWAPs when routing to the specific topology of the physical qubits. This SWAP-routing problem is NP-hard \cite{tan2020optimality}, and the Qiskit transpiler---set to its maximum optimization level (\texttt{optimization\_level=3})---solves it heuristically using the SABRE-SWAP method from Ref. \cite{li2019tackling} (Algorithm 1).

There are typically a number of subgraphs of qubits on the physical hardware that satisfy the minimum number of SWAPs. Since no two qubits are identical in superconducting hardware, however, the best subgraph needs to be searched out. We call this process optimizing the qubit layout. To get the best performance possible, it is best to select a qubit layout with qubits that have the longest $T_1$ and $T_2$ times, lowest gate errors which include decoherence effects due to their finite duration, and lowest readout errors. It can sometimes be impossible to satisfy all of these criteria simultaneously, but estimating which qubits to use based on these factors is possible. A recently added open-source Qiskit extension called \texttt{mapomatic}~\cite{mapomatic} provides a quick estimate of which are the best qubits to use across many devices. Mapomatic uses the VF2 subgraph isomorphism algorithm~\cite{vf2-cordella} to find compatible layouts on a physical architecture and scores them with a heuristic calculated from the current calibration data. Currently, single and two-qubit gate errors as well as measurement errors are considered in the calculation, and users may input a cost function of their choosing.

For the tight-binding simulation, we have chosen the seven qubit device $\textit{ibm\_lagos}$. Since the geometry of the tight-binding model is a 1D chain of five sites with nearest neighbor interactions, it's best to choose a matching qubit layout---five qubits in a line---because that layout would not need SWAPs. This leaves four unique layouts as shown in Fig. \ref{fig:lagos-ql-choice}. At the time of execution, Fig. \ref{fig:lagos-ql-choice3} was the best qubit layout, as scored by \texttt{mapomatic}. 

\subsection{Reduce number and duration of two-qubit gates} For current quantum computers, the fidelity of two-qubit gate operations is often the limiting performance factor. Two-qubit gate operations are roughly an order of magnitude more prone to errors than single-qubit gates, 
so reducing their number can lead to significant performance improvements.
This can be quite dramatic assuming our quantum simulation circuits have a medium to long depth, though it ultimately depends on the specifics of the two-qubit gate fidelity and the number of two-qubit gates.

Reducing the depth, or the execution duration, of the circuit by transpilation is often a good starting point. In Qiskit, the transpiler is a broad term for a group of circuit transformations which are used to map an abstract circuit onto a physical backend. It takes a quantum circuit and represents it as a directed acyclic graph (DAG), performing analysis and transformation `passes' on the graph. A pass manager orchestrates theses passes and attempts to reduce the number (and/or cost) of gate operations in the circuit. There are many transpiler options, the highest of which is \texttt{optimization\_level=3} which increases optimization searching for shorter, equivalent gate sequences at the expense of longer transpilation time. Each optimization level uses a preconfigured set of passes. To implement a custom optimization strategy, a user can specify their own pass manager. We specify passes that collect gate operations into two-qubit unitaries called blocks. The two-qubit blocks are then decomposed into a minimal set of two-qubit gate operations using the KAK decomposition~\cite{tucci2005introduction, earnest2021pulse}, after which the single-qubit gate operations are reduced. See the source code Jupyter notebook for further details \cite{tutorial-notebook}.

\begin{figure}[t]
  \centering   
  \includegraphics[width=0.9\linewidth]{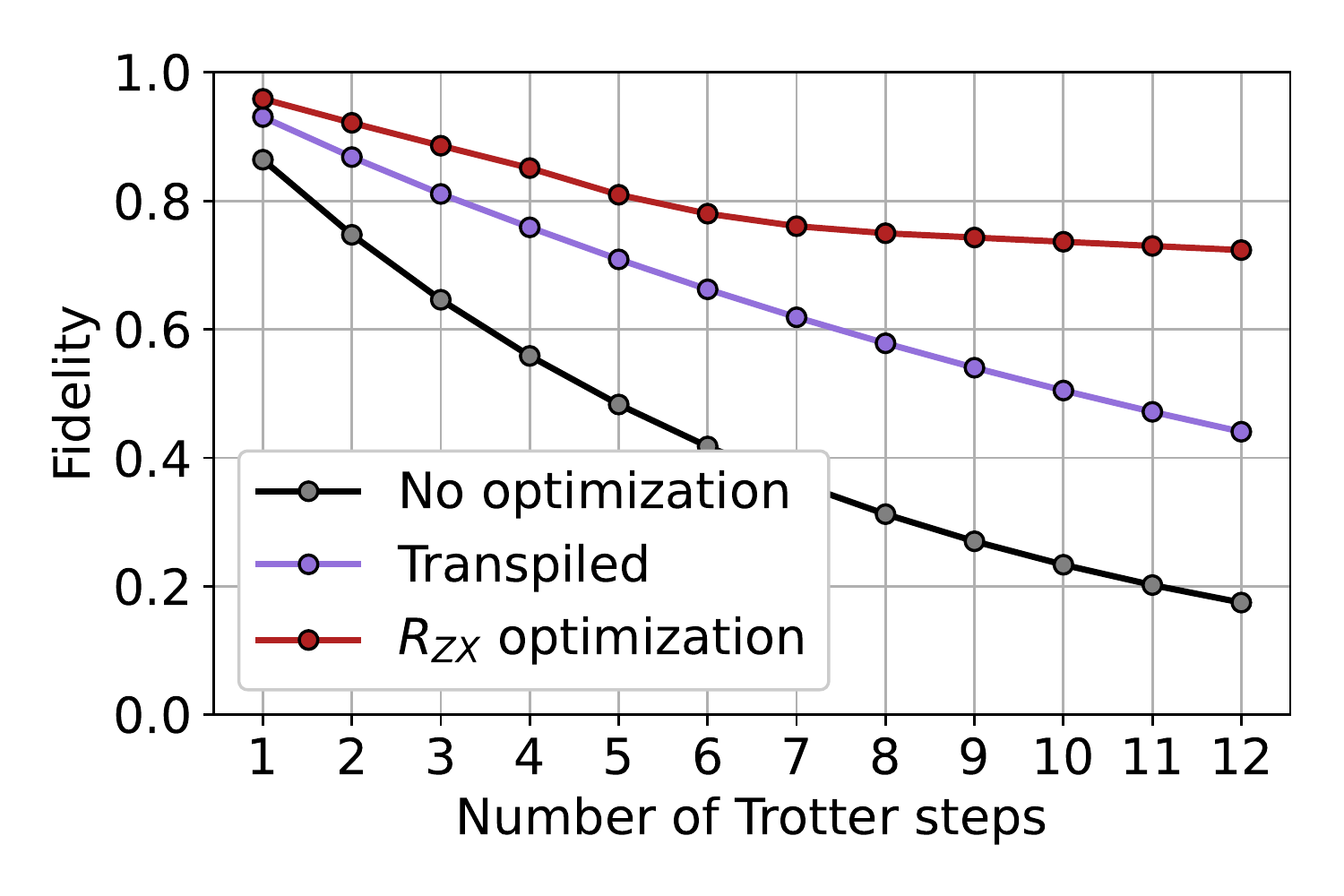} 
  \caption{Estimate of the circuit fidelity for circuits of various Trotter steps $m$ and fixed evolution time. With no optimization (black) of the quantum circuit the fidelity drops dramatically for larger Trotter steps. In purple is the circuit fidelity after heavily optimizing the circuit using Qiskit's transpiler as a black box optimization. The improvement comes mostly from the reduced number of two-qubit gates. The $R_{ZX}$ optimized circuit (red) fidelity is averaged over the same time range used in the quantum simulation dynamics of Fig. \ref{fig:trotter-error} $[0, 4 \pi]$.}
  \label{fig:trotter-steps-vs-fid}
\end{figure}

Another, and often better, strategy to reduce gate number and gate duration to compose quantum circuits in terms of two-qubit gates that are native to the device instead of the more common two-qubit gate operations such as the CNOT \cite{Stenger2021,earnest2021pulse}. For example, just to construct a CNOT, IBM Quantum hardware has historically used the echoed cross-resonance \cite{sheldon2016procedure, chow2011simple}. This effectively generates the interaction $R_{ZX}(\theta) = e^{-i\frac{\theta}{2} (X \otimes Z)}$ between two qubits and by wrapping $R_{ZX}(\pi/2)$ with some single qubit rotations, produces a CNOT gate operation. If instead, the circuit is originally optimized for $R_{ZX}(\theta)$, the circuit will typically need fewer $R_{ZX}(\theta)$s and with shorter duration ($\theta < \pi/2$)---as will be demonstrated below.

\begin{figure}[t]
  \centering   
  \includegraphics[width=0.9\linewidth]{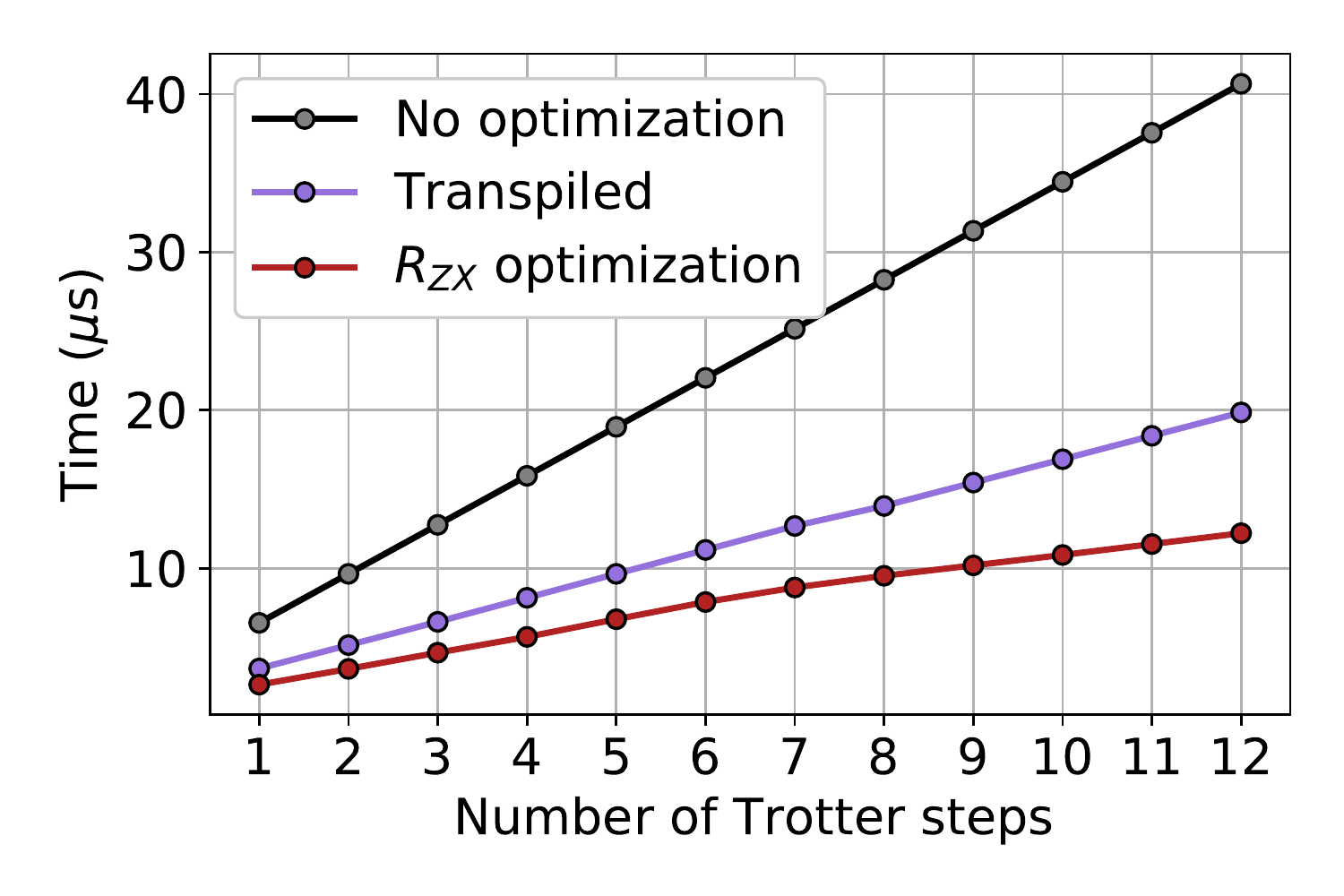} 
  \caption{Estimate of the total runtime on $\textit{ibm\_lagos}$ for circuits of increasing Trotter steps $m$. The circuit without any optimization is shown in black. In purple is a transpiled circuit using the Qiskit transpiler but treated as a black box. In red is the transpiled circuit with the additional switch from CNOT gate operations to native $R_{ZX}(\theta)$ gates.}
  \label{fig:trotter-steps-vs-dur}
\end{figure}

We construct a simple heuristic to estimate the fidelity of a circuit and compare the transpiler optimization against using the native $R_{ZX}(\theta)$ gate operation for various Trotter steps $m$ Fig. \ref{fig:trotter-steps-vs-fid}. For $N_{\text{1Q}}$ single-qubit gate operations and $N_{\text{2Q}}$ two-qubit gate operations the circuit fidelity is estimated by
\begin{equation}
    f_{\text{circuit}} = \bar{f}_{\text{1Q}}^{N_{1Q}} \bar{f}_{\text{2Q}}^{N_{2Q}} \;.
    \label{eq:qc-fid}
\end{equation}
At the time of execution, the average two-qubit gate fidelity for $\textit{ibm\_lagos}$ was $\bar{f}_{\text{2Q}} = 99.4\%$ and the average single-qubit gate fidelity was $\bar{f}_{\text{1Q}} = 99.98\%$. We also convert the quantum circuits into a Qiskit Pulse schedule which contains more detailed hardware control information such as an estimate of the circuit's runtime on $\textit{ibm\_lagos}$ (Fig. \ref{fig:trotter-steps-vs-dur}). The runtime information in conjunction with the qubits' $T_1$ and $T_2$ times can estimate errors from relaxation ($T_1$) and dephasing ($T_2$). The circuit runtimes are short compared to the average $T_1$ and $T_2$ times for $\textit{ibm\_lagos}$, so we will focus our discussion on the circuit fidelity.

For a short circuit, such as a single Trotter step, the total circuit fidelity improves from no optimization at $86.4\%$ to $93.0\%$ after transpilation. For a longer circuit such as 10 Trotter steps, however, the fidelity \textit{doubles} from $23.0\%$ to $50.5\%$ after transpilation as well as shortening the runtime by half. Overall, the transpiler improves the circuit performance significantly, however, much more can be done to improve the fidelity.

Assuming a linear dependence between the gate error and the angle of rotation $\theta$ for an $R_{ZX}(\theta)$ gate operation \cite{Stenger2021}, the circuit fidelity for the $R_{ZX}(\theta)$ optimization (red) is higher for all $m$ and almost flattens out for an increasing number of Trotter steps. To understand this, consider that as the number of Trotter steps $m$ increases, more single-qubit gate operations and $R_{ZX}(\theta)$ gate operations are in the circuit which leads to lower fidelity. However, the sum of the $R_{ZX}$ rotation angles remains fixed since the total evolution time is fixed for the various $m$ Trotter steps in these plots. This means the error contribution from two-qubit gate operations remains fixed for the $R_{ZX}(\theta)$ optimization---under these assumptions---and gives a somewhat flattened fidelity as $m$ increases. The more shallow decrease in fidelity after $m=7$ is mostly driven by single-qubit gate errors.


\subsection{Example $R_{ZX}$ decomposition}
\begin{figure}[t]
    \begin{quantikz}[column sep=0.25cm, row sep=0.25cm]
    \qw &\gate{H} & \ctrl{1} & \qw & \ctrl{1} & \gate{H} & \midstick[2,brackets=none]{=} \qw & \gate{H} & \gate[2]{R_{ZX}(\theta)} & \gate{H} & \qw \\
    \qw & \gate{H} & \targ{} & \gate{R_Z(\theta)} & \targ{} & \gate{H} & \qw & \qw & \qw & \qw & \qw \\ 
    \end{quantikz}
    \caption{A circuit equivalence for the $R_{XX}(\theta)$ two-qubit gate for more efficient implementation on superconducting qubits. Instead of two CNOT gates, a single and native two-qubit $R_{ZX}(\theta)$ interaction can be generated by echoed cross resonance. $R_{ZZ}(\theta)$ and $R_{YY}(\theta)$ gate operations can be implemented similarly by using different single-qubit gate operations.}
    \label{fig:rzx-gate}
\end{figure}
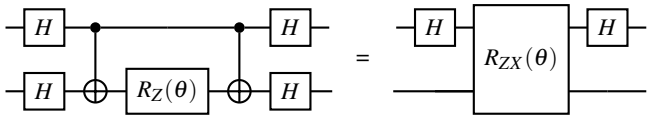

Consider Fig. \ref{fig:rzx-gate} where the standard decomposition of $R_{XX}(\theta)$ requires two CNOT gates. On IBM backends, each CNOT gate is constructed using cross-resonance pulses that generate fully entangling operations $R_{ZX}(\pi/2)$ in addition to a few single-qubit gates. This requires two $R_{ZX}(\pi/2)$ gate operations to execute the $R_{XX}(\theta)$ gate from Fig. \ref{fig:rzx-gate}. Alternately---and often with significantly less error---the $R_{XX}(\theta)$, $R_{YY}(\theta)$, or $R_{ZZ}(\theta)$ gate operations can be directly constructed from a partially entangling cross-resonance pulse $R_{ZX}(\theta)$ with some single-qubit gate operations before and after to transform the $R_{ZX}(\theta)$ interaction into the desired two-qubit rotation (see Fig. \ref{fig:rzx-gate}). By partially entangling, we mean the angle of rotation $\theta \neq \pi/2$. Applying this $ R_{ZX}(\theta)$ optimization, the quantum device would execute half as many $R_{ZX}$ gate operations and the average rotation angle per gate is usually less than $\pi/2$ due to the smaller angles required by Trotterization. Hence for the Trotterized unitary time evolution operator (Eq. \eqref{eq:tb-u-trotterized}), we compose each $R_{XX}$ and $R_{YY}$ gate from an echoed $R_{ZX}$ gate scaled to produce a rotation of $\theta$ in the two-qubit Hilbert space with the addition of appropriate single-qubit gates.

\subsection{Measurement error mitigation}
Measurement error mitigation reduces errors related to reading out qubit states. A simple, yet powerful, approach is to characterize each qubit's state preparation and measurement (SPAM) errors through a series of experiments. Those results are then used to counteract the SPAM errors of a future experiment through a post-processing analysis.

Consider a vector $\vec{p} = \{p(0), p(1)\}$ of the ideal probabilities measured by a noiseless quantum computer where $p(0)$ is the probability of measuring the state $|0\rangle$ and $p(1)$ is the probability of measuring $|1\rangle$. Now consider the actual probability vector $\vec{p_n}$ returned by a real (and noisy) quantum computer. Treating the errors classically, the confusion matrix $A$ can relate the ideal and noisy probabilities vectors $\vec{p_n} = A \vec{p}$. The matrix $A$ can be measured with simple state preparation circuits, such as the preparation and measurement of every possible bit-string outcome. After the elements of $A$ are known, the ideal probability vector of any measurement can be reconstructed from the noisy output $\vec{p_n}$ by applying $A^{-1}$ to the noisy data: $\vec{p} = A^{-1} \vec{p_n}$. 

The prescribed method above does not scale well, but improved methods have shown promise for scalability without loss of error suppression \cite{nation2021scalable}. There are two main scalability issues. First, the number of experiments needed to fully construct $A$ scales exponentially with the number of qubits ($2^N$) making a complete characterization of $A$ impractical. Second, classically computing the inverse of $A$ will be costly for many qubits. There are, however, various alternative or approximate methods which yield nearly identical results to a full characterization. They do not require exponentially many experiments nor do they require matrix inversion \cite{bravyi2021mitigating, nation2021scalable} (and can be implemented using the \texttt{mthree} Qiskit extension\cite{mthree}) .

We demonstrate a full characterization of SPAM errors on the five mapomatic selected qubits in the Jupyter notebook. As an example for how this can improve performance, we generate a five qubit GHZ state $(|00000\rangle + |11111\rangle) / \sqrt{2}$. The probability amplitudes, without measurement mitigation, are measured to be $44.7\pm 0.5\%$ for $|00000\rangle$ and $45.4\pm 0.5\%$ for $|11111\rangle$. After applying measurement mitigation in post-processing, the probability amplitudes improve to $48.7\pm 0.5\%$ for $|00000\rangle$ and $47.7\pm 0.5\%$ for $|11111\rangle$ (as opposed to their ideal amplitude of 50\% for each state). This is a notable improvement of $8.9\%$ and $5.2\%$ respectively.

\subsection{Post-selection based on physical constraints}
The last method discussed in this article is to use physical constrains or symmetries of the simulated system to mitigate errors. In the case of the tight-binding model, there is a constraint on particle number. Since particle number is conserved in many fermionic systems such as this model, the simulation must end with the same number of particles it started with. Hence, the quantum computer should only ever output bitstrings (the measured states) that have one 1 and four 0s. For example, if the quantum computer returns the state $|00000\rangle$ the particle was ``destroyed.'' If $|11000\rangle$ is measured, a second particle was ``created.'' In both cases, the Hamiltonian we are simulating could not have done this, so we can assume there was an error. Such measurements are ignored in post-processing. By particle conservation and starting with a single particle, the states $|10000\rangle$, $|01000\rangle$, $|00100\rangle$, $|00010\rangle$, or $|00001\rangle$ should be the only measured states. This is a powerful error mitigation option, but comes at the cost of potentially throwing out many measurements as well as being specific to the system and observable at hand \cite{Sung2022}.

\subsection{Results}


\begin{figure}
    \centering
    \includegraphics[width=\linewidth]{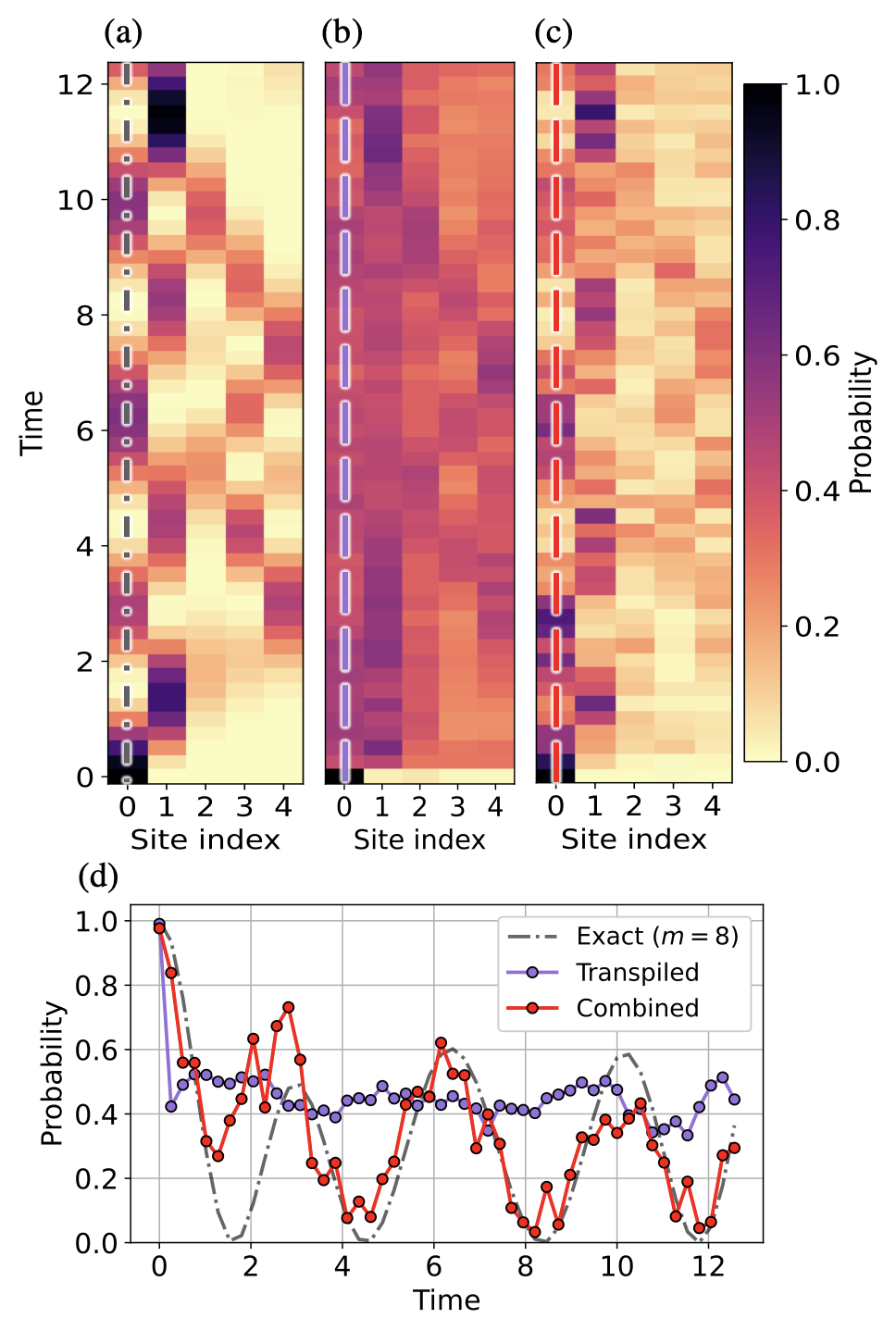}
    \caption{Trotterized time evolution of the initial state $|\psi_0\rangle = |00001\rangle$ for $m=8$ Trotter steps. Panel (a) is the exact calculation. Panel (b) is the state evolution computed by $\textit{ibm\_lagos}$ after transpilation. Panel (c) is the state evolution computed by $\textit{ibm\_lagos}$ when combining all the error mitigation techniques discussed in this section. The vertical lines along site 0 in Panels (a-c) show the location of the data shown in Panel (d).}
    \label{fig:evo_camp}
\end{figure}

On the quantum computer $\textit{ibm\_lagos}$, we examine the time evolution of the inital state $\ket{\psi_0} = \ket{00001}$ with $m=8$ Trotter steps for two different experimental optimizations (Fig. \ref{fig:evo_camp}(b-c)) and compare them to the exact unitary evolution (Fig. \ref{fig:evo_camp}(a)). In practice, we will want to choose $m$ such that the Trotter error and hardware error are both minimized. Here, we've selected $m=8$ to illustrate the significant improvements achievable with error mitigation.

The first optimization we will refer to as a ``transpiled'' optimization. The quantum circuit from Eq. \eqref{eq:tb-u-trotterized} is programmed as a quantum circuit in Qiskit and transpiled using Qiskit's transpiler without including any of the other optimization strategies mentioned in this section. This can be thought of as a blind, or black-box, optimization. From Figs. \ref{fig:trotter-steps-vs-fid} and \ref{fig:trotter-steps-vs-dur}, we know the transpiler will give significant improvement by reducing the circuit time and improve circuit fidelity by reducing the number of CNOT gates. However, the quantum state dynamics are quickly washed out by noise with no significant oscillation present (Fig. \ref{fig:evo_camp}(b,d)) . By delving deeper into quantum error mitigation and considering the physics of the system at hand as discussed previously, significant improvements can be achieved.

The second optimization, which we will refer to as the ``combined'' optimization, uses all the error mitigation methods mentioned in this section. The best qubit layout is selected using the \texttt{mapomatic} package. The full confusion matrix is measured and then used to correct for SPAM errors. The quantum circuit for $m=8$ Trotter steps is rewritten in terms of the device's native two-qubit interactions: $R_{ZX}(\theta)$ (pulse) gate operations. Specific transpiler passes are used to minimize the number of two-qubit and single-qubit gate operations which reduces circuit depth and circuit duration. We execute the circuit $10,000$ times to reduce shot noise, and in a post-selection process, measurements that do not preserve particle number are excluded. With this holistic approach, the combined optimization in Fig. \ref{fig:evo_camp}(c,d) shows much clearer dynamics overall with good agreement to the expected dynamics for $m=8$ Trotter steps Fig. \ref{fig:evo_camp}(a,d).  We also note that many of these utilities for optimizing the time evolution circuits can be found in this article's associated github repository.



\section{conclusion} \label{sec:conclusion}
Quantum simulation on today's quantum processors can seem like an formidable task, as there is no denying the current level of noise makes running these experiments challenging. However, there is still much to be gained from learning how to work around this noise and push the hardware to its limits. As the field moves forward, there are emerging trends in theory and experiment that are becoming standard practice.

This article has laid out a path for best practices in simulating quantum dynamics (Fig. \ref{fig:flowchart}). It provides brief, yet representative, introductions with more detailed discussion in the appendix. It then focused on specific techniques to realize the state evolution of a simple tight-binding model and combined multiple error mitigating techniques to produce powerful results despite initially high error estimates. 

We invite readers to extend and build off this work by considering different Hamiltonians, lattice connectivities, initial states, observables, measurement methods, and error mitigation techniques. There remain many scientific hurdles to achieve fault-tolerant quantum computation, but by applying state-of-the-art techniques, quantum simulation can be used to bring to light meaningful data and discoveries about fundamental physics. 

\section*{Acknowledgments}
The authors acknowledge the use of IBM Quantum Services for this work. Figures \ref{fig:trotter-steps-vs-fid}, \ref{fig:trotter-steps-vs-dur}, \ref{fig:trotter-error}, and \ref{fig:evo_camp} were generated from \texttt{matplotlib} \cite{Hunter:2007}. The authors would also like to acknowledge Thomas Alexander, Max Rossmannek, Kevin Sung, and Mirko Amico for useful discussions and source code.

\bibliographystyle{unsrt}
\bibliography{refs} 

\clearpage

\appendix
\section{Fermionic State Preparation}
\label{appendix:state-prep}
In some cases, one may wish to prepare an eigenstate (often the ground state) of the target Hamiltonian as the initial state to evolve.  If this Hamiltonian conserves both number and parity, we can prepare  Slater determinants---eigenstates of Hamiltonians quadratic in fermionic second quantization operators.  These eigenstates can be prepared using a series of elementary operations known as Givens rotations which have the form
\begin{equation}
    G\left(\theta, \phi\right) = \begin{pmatrix}
        \cos\theta & -e^{i\phi}\sin\theta \\
        \sin\theta & e^{i\phi}\cos\theta
    \end{pmatrix}.
    \label{eq:Givens-rot}
\end{equation}
The preparation of a Slater determinant using Givens rotations, introduced in~\cite{Wecker-givensrotation}, proceeds in the following way.  In second quantization the Slater determinant is described as
\begin{equation}
        \ket{\Psi_{\mathrm{SD}}} = \prod_{i=1}^n b_j^\dagger\ket{\mathrm{vac}}\ \ \ \ \ b_j^\dagger = \sum_{i=1}^N a_i^\dagger B_{ij}
\end{equation}
where $\ket{\mathrm{vac}}$ is the vaccuum state, $a_i^\dagger$ are the fermionic creation operators, $n$ is the number of particles, $N$ is the number of orbitals in the system, and $B$ is an $n\times N$ matrix satisfying
\begin{equation}
    B^\dagger B = P_s
\end{equation}
where $P_s$ is the projector onto the subspace spanned by the single-particle wave functions of the occupied spin orbitals.  

\begin{figure}
    \centering
    \begin{quantikz}[column sep=0.35cm, row sep=0.4cm]
    \qw & \targ{} & \qw & \ctrl{1} & \qw & \targ{} & \qw & \qw\\
    \qw & \ctrl{-1} &\qw & \gate[]{e^{i\theta Y}} & \qw & \ctrl{-1} & \gate[]{e^{-i\phi Z/2}} & \qw
    \end{quantikz}
    \caption{Quantum circuit of the Givens rotation $\mathcal{G}\left(\theta, \phi\right)$ between two adjacent spin orbitals in the JWT ordering.}
    \label{fig:givens-circ}
\end{figure}
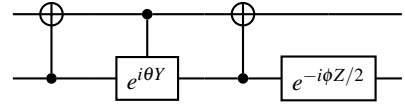

At this point, we can take advantage of the fact that an arbitrary Slater determinant can be prepared by applying a single-particle basis change $U$ to an easy-to-prepare determinant in the computational basis
\begin{equation}
        \ket{\Psi_{\mathrm{SD}}} = U\prod_{i=1}^n c_j^\dagger\ket{\mathrm{vac}}\ \ \ \ \ Uc_j^\dagger U^\dagger = b_j^\dagger.
\end{equation}
This unitary $U$ corresponds to a product of single-particle basis transformations and can be decomposed into a sequence of Givens rotations
\begin{equation}
    U = \mathcal{G}_1\mathcal{G}_2\cdots \mathcal{G}_{N_g}
\end{equation}
with each Givens rotation $\mathcal{G}(\theta, \phi)$ acting on the $j$th and $k$th orbitals as
\begin{equation}
    \begin{pmatrix}
        \mathcal{G}c_j^\dagger \mathcal{G}^\dagger \\
        \mathcal{G}c_k^\dagger\mathcal{G}
    \end{pmatrix} = G\left(\theta, \phi\right)\begin{pmatrix}
        c_j^\dagger \\ c_k^\dagger
    \end{pmatrix}
\end{equation}
where $G\left(\theta, \phi\right)$ is defined above in~\ref{eq:Givens-rot}.  Givens rotations of the form above can be implemented on a quantum computer using the circuit in Fig.~\ref{fig:givens-circ}.
To find the appropriate angles of each Givens rotation, one can utilize the open-source project \texttt{OpenFermion}~\cite{OpenFermion}.  In it, the Givens rotation between each pair of orbitals is given as output, taking the matrix $B$ as input.

\section{Gaussian State Preparation}
\label{appendix:gaussian-state}
If we wish to prepare the ground state of a Hamiltonian where particle number is not a constraint, fermionic gaussian states can be used~\cite{Jian-gaussianStates, verstraete-ferminoicstates}.  Given a quadratic Hamiltonian of the form
\begin{equation}
    H = \sum_{i,j}^N M_{i,j}c_i^\dagger c_j + \frac{1}{2}\sum_{i,j}^N \left(P_{i,j}c_i^\dagger c_j + P_{i,j}^* c_j^\dagger c_i\right)
    \label{eq:general-quadratic}
\end{equation}
where $M=M^\dagger$ and $P=-P^T$ are complex matrices, we will consider the procedure to bring $H$ into its diagonal form
\begin{equation}
    H = \sum_{i=1}^N \epsilon_ib_i^\dagger b_i + c.
\end{equation}
These new fermionic operators, $b_i^\dagger$ and $b_i$, are linear combinations of $c_i^\dagger$ and $c_i$ 
\begin{equation}
    \begin{pmatrix}
        \mathbf{b}^\dagger \\ \mathbf{b}
    \end{pmatrix} = W\begin{pmatrix}
        \mathbf{c}^\dagger \\ \mathbf{c}
    \end{pmatrix} = \begin{pmatrix}
        \mathcal{W}\mathbf{c}^\dagger \mathcal{W}^\dagger \\
        \mathcal{W}\mathbf{c}\mathcal{W}^\dagger
    \end{pmatrix}
\end{equation}
where $\left(\mathbf{c}^\dagger \ \ \mathbf{c}\right)^T = \left(c_1^\dagger \ c_2^\dagger \cdots c_N^\dagger \ c_1\ c_2 \cdots c_N\right)^T$ and $\left(\mathbf{b}^\dagger \ \ \mathbf{b}\right)^T = \left(b_1^\dagger \ b_2^\dagger \cdots b_N^\dagger \ b_1\ b_2 \cdots b_N\right)^T$.  The matrix $\mathcal{W}$ is a fermionic gaussian unitary which performs the linear transformation $W$.  With this transformation, the ground state of the Hamiltonian~\ref{eq:general-quadratic} is
\begin{equation}
    \ket{\psi_0} = \mathcal{U}\ket{\mathrm{vac}} = \mathcal{W}\mathcal{V}\ket{\mathrm{vac}}
\end{equation}
The matrix $W$ has the block form
\begin{equation}
    W = \begin{pmatrix}
        W_1^* & W_2^* \\ W_1 & W_2
    \end{pmatrix}
\end{equation}
whose submatrices satisfy
\begin{align}
    W_1W_1^\dagger + W_2W_2^\dagger = \mathbb{1} \\
    W_1W_2^T + W_2W_q^T = \mathbb{0}
\end{align}
with $\mathbb{1}$ and $\mathbb{0}$ being the identity and zero matrix respectively.  The bottom half of the matrix $W_B = \begin{pmatrix}W_1 & W_2\end{pmatrix}$ uniquely determines the transformation $\mathcal{W}$ up to a global phase and can be constructed through the procedure specified in Appendix A of~\cite{Jian-gaussianStates}.  The goal of Gaussian state preparation is then to find the unitary matrix
\begin{equation}
    U = \mathcal{B}\mathcal{G_{N_G}}\cdots\mathcal{B}\mathcal{G}_3\mathcal{G}_2\mathcal{B}\mathcal{G}_1\mathcal{B}
\end{equation}
such that
\begin{equation}
    VW_BU^\dagger = \begin{pmatrix}
        \mathbb{0} & \mathbb{1}
    \end{pmatrix}
\end{equation}
where: $V$ is an arbitrary matrix, $\mathcal{B}$ is a representation of a particle-hole transformation
\begin{equation}
    \mathcal{B}=\mathcal{B}^\dagger = \begin{pmatrix}
        \mathbb{1}-u_Nu_N^T & u_Nu_N^T \\ 
        u_Nu_N^T & \mathbb{1}-u_Nu_N^T
    \end{pmatrix}
\end{equation}
where $u_N = \begin{pmatrix} 0 & 0 & \cdots & 1\end{pmatrix}$ is a unit vector of length $N$, and $\mathcal{G}$ are Givens rotations of the form
\begin{equation}
    \mathcal{G} = \begin{pmatrix}
        \cos\theta & -e^{i\phi}\sin\theta & 0 & 0 \\
        \sin\theta & e^{i\phi}\cos\theta & 0 & 0 \\
        0 & 0 & \cos\theta & -e^{i\phi}\sin\theta \\
        0 & 0 & \sin\theta & e^{i\phi}\cos\theta
    \end{pmatrix}.
\end{equation}
The number of Givens rotations and particle-hole transformations is
\begin{equation}
    N_G = (N-1)\frac{N}{2},\ \ \ N_B=N
\end{equation}
and has a depth of at most $2N-1$.  Once this matrix, $U$ has been found, it can be decomposed into a sequence of quantum gate operations accessible to a quantum computer.  This algorithm has been implemented in the OpenFermion project~\cite{OpenFermion} and is described in more detail in~\cite{Jian-gaussianStates}.

\section{Linear Combination of Unitaries}
\label{appendix:LCU}
Another method for approximating the time evolution operator $U(t)$ has been found through the method of Linear Combination of Unitaries~\cite{Childs-LCU} (LCU).  This approach has the advantage of a higher degree of accuracy to the Trotter-Suzuki formalism at the cost of adding an auxiliary register of qubits.  We begin this procedure by considering the Taylor expansion of the time evolution operator (and borrowing the notation from Ref.~\cite{Somma-LCU})
\begin{equation}
    U(t) = e^{-iHt} \approx \sum_{k=0}^K\frac{\left(-it\right)^k}{k!}H^k
\end{equation}
where the series is truncated at order $K$.  If we write $H$ as a sum of interactions $H=\sum_i a_iH_i$, the above can be written as
\begin{equation}
    U_{aprx}(t)= \sum_{k=0}^K\frac{\left(-it\right)^k}{k!}\sum_{i_1,\cdots,i_k=1}^N a_{i_1}\cdots a_{i_k} H_{i_1}\cdots H_{i_k} 
\end{equation}
which is of the form
\begin{equation}
    \Tilde{U} = \sum_{j=0}^K \alpha_j V_j
\end{equation}
where $\alpha_k \geq 0$ and $V_k$ is a unitary corresponding to $(-i)^kH_{i_1}\cdots H_{i_k}$.

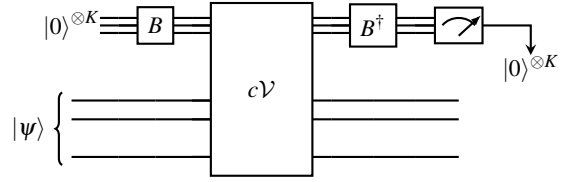
\begin{figure}
    \centering
    \begin{quantikz}[column sep=0.25cm, row sep=0.25cm]
    \ket{0}^{\otimes K} &\qwbundle[alternate]{}  & \gate{B} \qwbundle[
alternate]{}  & \qwbundle[alternate]{}& \gate[wires=4]{c\mathcal{V}}[2cm] \qwbundle[alternate]{} & \qwbundle[alternate]{} & \gate{B^\dagger} \qwbundle[
alternate]{} & \qwbundle[alternate]{} & \meter{}\qwbundle[
alternate]{} &  \trash{\text{$|0\rangle^{\otimes K}$}}
 \\
    \lstick[wires=3]{$\ket{\psi}$} &\qw &\qw &\qw  &\qw & \qw & \qw & \qw & \qw \\
    \qw & \qw & \qw & \qw & \qw & \qw & \qw & \qw & \qw  \\
    \qw & \qw & \qw & \qw & \qw & \qw & \qw & \qw & \qw  
    \end{quantikz}
    \caption{Circuit schematic of the method of the LCU method.  The auxiliary register is prepared in the state $\frac{1}{\sqrt{s}}\sum_{j=0}^{m-1}\sqrt{\alpha_j}\ket{j}$ via the multi-qubit operator $B$ before applying the unitary $c\mathcal{V}$.  To arrive at the final state $PW\ket{0}\ket{\psi} = \frac{1}{s}\ket{0}^{\otimes K}\tilde{U}\ket{\psi}$ we apply $B^\dagger$ on the auxiliary register and keep only states in which the auxiliary register was measured to be $\ket{0}^{\otimes K}$.}
    \label{fig:lcu-circuit}
\end{figure}

To perform the operator $\Tilde{U}$ on a state $\ket{\psi}$, we need a $K$-dimensional auxiliary register prepared in the state
\begin{equation}
    B\ket{0} = \frac{1}{\sqrt{s}}\sum_{j=0}^K\sqrt{\alpha_j}\ket{j}
\end{equation}
where $s\equiv \sum_j^K \alpha_j$.  Then suppose we can create a controlled unitary $c\mathcal{V}$ such that for each $j$
\begin{equation}
    c\mathcal{V}\ket{j}\ket{\psi} = \ket{j}V_j\ket{\psi}.
\end{equation}
Then defining
\begin{equation}
    W\equiv (B^\dagger \otimes \mathbb{1})c\mathcal{V}(B\otimes \mathbb{1}),
\end{equation}
the action of this on the state $\ket{0}^{\otimes K}\ket{\psi}$ yields
\begin{equation}
    W\ket{0}^{\otimes K}\ket{\psi} = \frac{1}{s}\ket{0}^{\otimes K}\Tilde{U}\ket{\psi} + \sqrt{1 - \frac{1}{s^2}}\ket{\Phi}
\end{equation}
where the auxiliary part of $\ket{\Phi}$ has support orthogonal to $\ket{0}^{\otimes k}$. Applying the projector $P=\ket{0}^{\otimes k}\bra{0}^{\otimes k}\otimes\mathbb{1}$ onto the state $W\ket{0}^{\otimes K}\ket{\psi}$ obtains $PW\ket{0}^{\otimes K}\ket{\psi} = \frac{1}{s}\ket{0}^{\otimes K}\Tilde{U}\ket{\psi}$.  A simple schematic exampe of this circuit is shown in Figure~\ref{fig:lcu-circuit}.  It can be shown that if $s=2$ and $\tilde{U}$ is unitary, one can utilize oblivious amplitude amplification to perform $\tilde{U}$ exactly~\cite{Somma-OAA}.  

\captionsetup[subfigure]{position=top}


\section{Auxiliary Measurement}
\label{appendix:aux-measurement}
\begin{figure}
    \centering
    \begin{subfigure}[b]{0.5\textwidth}
    \caption{}
    \begin{quantikz}[column sep=0.25cm, row sep=0.25cm]
    \ket{+} & \ctrl{1} & \qw & \octrl{1} & \meter{} & \left<2\sigma_+\right>\\
      \lstick[wires=3]{$\ket{\psi(t)}$}  & \gate[wires=3]{V} & \qw & \gate[wires=3]{U} & \qw \\
        & \qw &\qw & \qw & \qw\\
        \qw &\qw & \qw & \qw & \qw
    \end{quantikz}
    \label{fig:uv-measurement}
    \end{subfigure}
    \centering
    \begin{subfigure}[b]{0.5\textwidth}
    \caption{}
    \begin{quantikz}[column sep=0.25cm, row sep=0.25cm]
    \ket{+} & \ctrl{1} & \qw & \octrl{1} & \meter{} & \left<2\sigma_+\right>\\
      \lstick[wires=3]{$\ket{\psi}$}  & \gate[wires=3]{B} & \gate[wires=3]{U(t)} & \gate[wires=3]{A} & \qw \\
        & \qw &\qw & \qw & \qw\\
        \qw &\qw & \qw & \qw & \qw
    \end{quantikz}
    \label{fig:correl-measurement}
    \end{subfigure}
    \caption{(a) A circuit measuring an observable of the form $\left<U^\dagger V\right>$ using an auxiliary qubit.  The auxiliary qubit is prepared in the $\ket{+} = (\ket{0} + \ket{1})/\sqrt{2}$ state and measures $\left<2\sigma_+\right> = \left<X + iY\right>$. (b) Circuit measuring the time correlation function of the two observables $\mathcal{C}_{AB}(t) = \left<A^\dagger(t)B(0)\right>$.  The black dot represents a $\ket{1}$-controlled gate and the white a $\ket{0}$-controlled gate.}
    \label{fig:aux-measurement}
\end{figure}
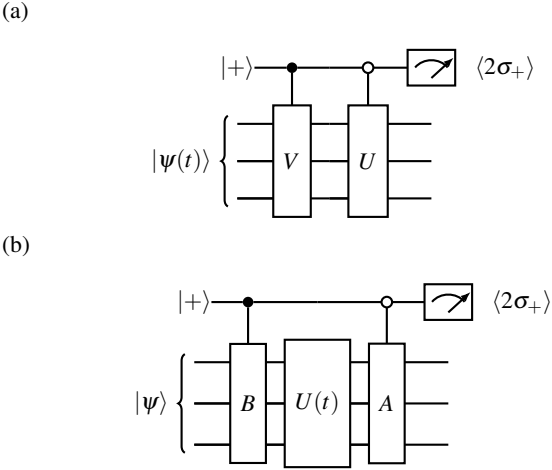

Additionally, observables, such as time-correlation functions, can be measured using an auxiliary qubit if the observable has the form $\left<U^\dagger V\right>$ using the circuit shown in Fig.~\ref{fig:uv-measurement}~\cite{Somma2002}.  To demonstrate this, consider the total state $\Tilde{\psi}$ after the circuit 
\begin{equation}
    \ket{\Tilde{\psi}} = \frac{1}{\sqrt{2}}\left(\ket{0}U\ket{\psi(t)} + \ket{1}V\ket{\psi(t)}\right)
\end{equation}
and the measurement
\begin{equation}
    \left<X + iY\right> = \left<HZH\right> + \left<ZHZH\right>.
    \label{eq:aux-measurement}
\end{equation}
Measuring the first term of~\ref{eq:aux-measurement} yields
\begin{align}
    \bra{\Tilde{\psi}}HZH\ket{\Tilde{\psi}} =& \frac{1}{2}\left(\bra{0}U^\dagger\bra{\psi(t)} + \bra{1}V^\dagger\bra{\psi(t)}\right)\nonumber \\
    &\cdot\left(\ket{0}V\ket{\psi(t)} + \ket{1}U\ket{\psi(t)}\right) \nonumber \\
    = & \frac{1}{2}\bra{\psi(t)}U^\dagger V\ket{\psi(t)}+ \frac{1}{2}\bra{\psi(t)}V^\dagger U \ket{\psi(t)}
\end{align}
and measuring the second term similarly gives
\begin{align}
    \bra{\Tilde{\psi}}ZHZH\ket{\Tilde{\psi}} =& \frac{1}{2}\left(\bra{0}U^\dagger\bra{\psi(t)} + \bra{1}V^\dagger\bra{\psi(t)}\right)\nonumber \\
    &\cdot\left(\ket{0}V\ket{\psi(t)} - \ket{1}U\ket{\psi(t)}\right) \nonumber \\
    = & \frac{1}{2}\bra{\psi(t)}U^\dagger V\ket{\psi(t)}- \frac{1}{2}\bra{\psi(t)}V^\dagger U \ket{\psi(t)}.
\end{align}
In total, the above equations show that
\begin{equation}
    \bra{\Tilde{\psi}}X + iY\ket{\Tilde{\psi}} = \bra{\psi(t)}U^\dagger V \ket{\psi(t)}.
\end{equation}

\captionsetup[subfigure]{position=top}

Auxiliary qubit measurements are also useful for observables such as correlation functions of the form~\cite{Ortiz2001} 
\begin{equation}
    \mathcal{C}_{AB}(t) = \left<A^\dagger(t)B(0)\right> = \left<e^{iHt}A^\dagger e^{-iHt} B\right>.
\end{equation}
The circuit shown in Fig.~\ref{fig:correl-measurement} describes the measurement of $\mathcal{C}_{AB}(t)$.  Similar to the example above, the state of the circuit just before measurement is
\begin{equation}
    \ket{\Tilde{\psi}} =  \frac{1}{\sqrt{2}}\left(\ket{0}AU(t)\ket{\psi(t)} + \ket{1}U(t)B\ket{\psi(t)}\right)
\end{equation}
and the measurement of $\left<X\right>_a$ on the auxiliary qubit returns
\begin{align}
    \left<X\right>_a =& {\rm Tr}\left[\left(X_a \otimes \mathbb{1}\right)\ket{\Tilde{\psi}}\bra{\Tilde{\psi}}\right] \nonumber \\
    =& {\rm Re}\left[\mathcal{C}_{AB}(t)\right]
\end{align}
the real part of $\mathcal{C}_{AB}(t)$.  The imaginary part of $\mathcal{C}_{AB}(t)$ is obtained in the same way with the measurement of $\left<iY\right>_a$.  This algorithm can also be generalized to extract $n$-point correlation functions as well as the expectation value of any operator with the form $\mathcal{Q} = \sum_i q_iU^\dagger_i V_i$ assuming $U_i$ and $V_i$ are unitary~\cite{Tacchino2020}.

\section{Eigenvalue Estimation}
\label{appendix:eig-estimation}
Additionally, the goal of quantum simulation is often to obtain the spectrum (eigenvalues) of a quantum system.  Generally, the time evolved wavefunction can be written as a linear superposition of the eigenvalues, $p_i$, and eigenvectors, $\ket{\varphi_i},$ of an operator $\mathcal{P}$ such that
\begin{equation}
    e^{-i\mathcal{P}t}\ket{\psi} = \sum_i c_i e^{-ip_i t}\ket{\varphi_i}.
\end{equation}
Finding the eigenvalues is then a matter of obtaining the phases, $p_i$, generally using Quantum Phase Estimation (QPE).  The standard QPE algorithm utilizes many controlled $\mathcal{P}^k$ gates which are then fed into a Quantum Fourier Transform(QFT)~\cite{nielsen-chuang, cleve-QFT}, making it prohibitively expensive in both circuit depth and the number of qubits required (though other methods such as the iterative phase estimation~\cite{Corcoles2021} have been shown to reduce the resources required).  

To avoid this, we can instead use a classical Fast Fourier Transform (FFT) on the expectation value $\bra{\psi}e^{-i\mathcal{P}t}\ket{\psi}$ which is measured using auxiliary qubits as in Appendix~\ref{appendix:aux-measurement}. Alternatively, the classical FFT can be computed directly on the state dynamics without coupling to an auxiliary qubit to extract the spectrum of $\mathcal{P}$ \cite{richerme2022}. If we set $V=e^{-i\mathcal{P}t}$ and $U=\mathbb{1}$ in the circuit of Fig.~\ref{fig:uv-measurement}, the measurement of the auxiliary qubit is
\begin{equation}
    \left<e^{-i\mathcal{P}t}\right> = \sum_i |c_i|^2e^{-ip_it}
\end{equation}
whose FFT yields
\begin{equation}
    FFT\left(\left<e^{-i\mathcal{P}t}\right>\right) = \sum_i 2\pi |c_i|^2\delta\left(p-p_i\right)
\end{equation}
the spectrum of $\mathcal{P}$.  The caveat to this approach is that we must be able to apply the time evolution operator for very long times in order to obtain the full spectrum.  To mitigate this, algorithms using a ``binning'' approach to obtain the eigenvalues have been proposed~\cite{Somma2019} which reduce the maximum simulation time at the expense of the accuracy and resolution of $p_i$.

\section{Spectroscopic Eigenvalue Estimation}
\label{appendix:spectroscopic-measurement}
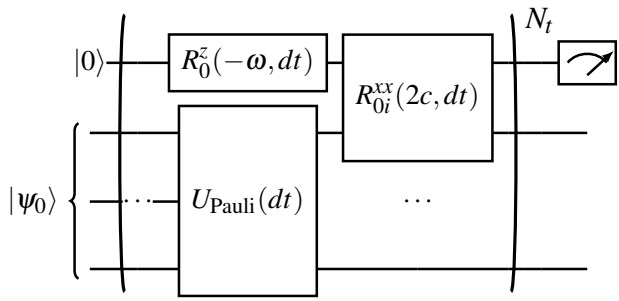
\begin{figure}[h]
\centering
\begin{tikzpicture}

\node[scale=1.2, anchor=south west] (circuit) at (0,0) {
    \begin{quantikz}[column sep=0.18cm, row sep=0.15cm]
      \ket{0} & \qw & \gate{R_0^z(-\omega,dt)} & \gate[wires=2]{R_{0i}^{xx}(2c,dt)} & \qw & \qw & \qw & \meter{} \\
      \lstick[wires=3]{$\ket{\psi_0}$} & \qw & \gate[wires=3]{U_{\rm Pauli}(dt)} &\ghost{R_{0i}^{xx}(2c,dt)} & \qw & \qw & \qw & \qw \\
        & \qw\cdots & \ghost{U_{\rm Pauli}(dt)} & \cdots & & & & \\
      & \qw & \ghost{U_{\rm Pauli}(dt)} & \qw & \qw & \qw & \qw & \qw
    \end{quantikz}
};
\begin{scope}[x={(circuit.south east)},y={(circuit.north west)}]
    \draw[decorate, decoration={bent,aspect=0.005},  very thick,rounded corners] (0.22, 0.08) -- (0.22, 1.0);
    \draw[decorate, decoration={bent,aspect=0.005},  very thick,rounded corners] (0.81, 1.0) -- (0.81,0.08);
    \node[scale=1.3] at (0.855,0.98) {$N_t$};
\end{scope}
\end{tikzpicture}
    \caption{Circuit diagram of the spectroscopic approach to estimating the spectrum of an operator $\mathcal{P}$.  Here $P = U_{\rm Pauli}$, $dt$ is the time step and $c$ is a coupling strength between the system and probe qubits.}
    \label{fig:spectroscopic-circ}
\end{figure}

An alternative, spectroscopic approach to estimating the spectrum of a Hamiltonian $H_{\rm Pauli}$ encoded to Pauli operators has been proposed~\cite{Stenger2022} which does not require the application of controlled time-evolution gates.  This approach is analogous to the experimental technique of spectroscopy, in which the response of a system is measured by sweeping the energy of a probe.  The circuit is shown in Fig.~\ref{fig:spectroscopic-circ} and illustrates the probe qubit ($q_0$ in this case) being pumped as a function of energy $\omega dt$ and interacting with a single qubit ($q_i$) of the system qubit register which has evolved to some time $dt$.  Subsequently measuring $\left<Z\right>$ on the probe qubit after $N_t$ applications of the pump/probe gates as a function of $\omega$ yields a spectrum of the simulated system, with dips corresponding to the eigenvalues of the operator encoding in $U_{\rm Pauli}=e^{idtH_{\rm Pauli}}$ 

\end{document}